\documentclass[journal]{IEEEtran}

\IEEEoverridecommandlockouts                              

\usepackage{amssymb}
\usepackage{amsmath}
\usepackage{bm}
\usepackage{mathrsfs}
\usepackage{comment} 
\usepackage{algorithmic,algorithm}
\usepackage{shuffle}
\usepackage{graphicx} 
\usepackage{caption}
\usepackage{lipsum}
\usepackage{color}
\usepackage{enumerate}
\usepackage{amsfonts}
\usepackage{subfigure} 
\usepackage{array}
\usepackage{setspace}
\usepackage{multirow}

\DeclareSymbolFont{symbolsC}{U}{pxsyc}{m}{n}
\DeclareMathSymbol{\coloneqq}{\mathrel}{symbolsC}{"42}

\newcommand{\tabincell}[2]{\begin{tabular}{@{}#1@{}}#2\end{tabular}}

\begin{document}

\title{\LARGE \bf
On Decidability of Existence of Fortified Supervisors Against Covert Actuator Attackers} 
\author{Ruochen Tai, Liyong Lin and Rong Su

%
%
%
\thanks{The research of the project was supported by the Agency for Science, Technology and Research (A*STAR) under its IAF-ICP Programme ICP1900093 and the Schaeffler Hub for Advanced Research at NTU.
The authors are affliated with Nanyang Technological University, Singapore. (Email: ruochen001@e.ntu.edu.sg; llin5@e.ntu.edu.sg; rsu@ntu.edu.sg). (\emph{Corresponding author: Liyong Lin})}
}
\maketitle

\begin{abstract}
This work investigates the problem of synthesizing fortified supervisors against covert actuator attackers. For a non-resilient supervisor $S$, i.e., there exists at least a covert actuator attacker that is capable of inflicting damage w.r.t $S$, a fortified supervisor $S'$ satisfies two requirements: 1) $S'$ is resilient against any covert actuator attacker, and 2) the original closed-behavior of the closed-loop system under $S$ is preserved, that is, $S'$ is control equivalent to $S$. By designing a sound and complete procedure, we show the problem of determining the existence of a fortified supervisor against covert actuator attackers is decidable. We also discuss how to extend the decidability result to the case against the worst-case attacker.
\end{abstract}




\section{Introduction}
\label{sec:intro}

The rapid development of cyber-physical systems (CPS) has made the network an indispensable element, but it is also vulnerable to compromise and exploitation by malicious attacks. As a result, there has been an increasing focus on resilient control to combat sophisticated cyber attacks in the control systems community. Attackers targeting CPS can generally be categorized as either covert or non-covert. In the case of non-covert attacks, detection mechanisms and isolating compromised components can be employed to counteract malicious actors effectively. However, the emergence of covert attacks poses a significant threat to CPS, as they can undermine system functionality stealthily without triggering conventional detection and localization methods. In real-world scenarios, covert cyber attacks have already led to serious incidents, such as the Stuxnet worm's attack on Iran's nuclear facilities \cite{FR11}, the Maroochy sewage control incident \cite{SM07}, and blackouts in the Ukrainian power grid \cite{LAC16}. Therefore, a crucial and challenging task is to ensure the secure operation of CPS by designing resilient controllers against covert attacks.

In the classical control community, abundant works have been devoted to resilient control against covert attacks (please refer to survey paper \cite{DZA22} for more details). However, there are few results (see \cite{Su2018}-\cite{Zhu2018}) on this topic at the supervisory control layer. 
\cite{Su2018}-\cite{LZS19b} synthesize resilient supervisors against covert sensor(-actuator) attackers. \cite{Zhu2018} studies supervisor fortification, which fortifies a supervisor to generate a new control equivalent supervisor resilient against covert actuator attacks. Nevertheless, only the work of \cite{Su20} provides a decidability result on the existence of resilient supervisors against covert sensor attacks. It is still an open question for the resilient control against covert actuator attacks. In fact, before tackling any complexity challenge to synthesize a resilient solution, it is essential to address a fundamental computability question regarding the existence of solutions. Thus, this paper will continue our previous work \cite{Zhu2018} and partially answer the above open question by showing that the existence of fortified supervisors against covert actuator attacks is decidable.

Although \cite{Su2018}-\cite{Zhu2018} are the most relevant to our work, there exist several differences. In terms of the problem setup, none of \cite{Su2018}-\cite{LZS19b} has studied the control equivalence. In addition, \cite{Su2018} and \cite{Su20} only consider sensor attacks that must have the same observations as the supervisor, different from our setup which considers actuator attacks that may have observations different from those of the supervisor. In terms of the technical methods, \cite{Su2018} proposed a heuristic approach, and \cite{LZS19b}-\cite{Zhu2018} proposed constraint-based approaches, but all of them are incomplete and cannot solve the decidability issue. \cite{Su20} solves the problem of determining the existence of a resilient supervisor against covert sensor attacks. However, due to the differences mentioned above, our setup is quite different from that of \cite{Su20}, resulting in that the technique of \cite{Su20} to prove the decidability fails for our problem. In addition, \cite{Su20} adopts transducer-based models and develops a language-based approach. In contrast, we adopt finite state automata for modeling and develop an automaton-based approach. Thus, our method is different from that of \cite{Su20}.

We remark that the supervisor fortification strategy is different from control strategies studied in \cite{GSS19}-\cite{WP19} because the fortified supervisor is defending against all possible \emph{covert} attacks while the supervisors considered in \cite{GSS19}-\cite{WP19} are defending against \emph{a given attack model}, e.g., the worst-case attack that may be non-covert. 
Compared with \cite{CWKL16}-\cite{GSLG19} that study diagnostic methods to disable controllable events after attack detection, the supervisor fortification focuses on supervisor synthesis without using a diagnostic tool. In addition, disabling all controllable events is not required in our work. There are some other defense strategies proposed in \cite{HMT21}-\cite{SFS21}, which are however different from the fortification strategy.

Note that determining the existence of A fortified supervisor against ALL covert actuator attacks is a non-trivial ``exist-for all'' computability question because of the following two challenges. Firstly, there can be an infinite number of control equivalent supervisors. Thus, exhaustively verifying the resilience of each control equivalent supervisor like \cite{Zhu2018} is infeasible. Secondly, there can be an infinite number of covert actuator attackers for any supervisor. Thus, the attack model is not fixed. Consequently, it is infeasible to reduce this problem to a standard supervisory control problem like \cite{GSS19}-\cite{MC2022indefinite}, which treat the composition of the plant and the given attacker as a new plant and the resilient supervisor to be synthesized as a new supervisor. 

To handle those challenges, we design a sound and complete procedure that terminates within finite steps, consisting of four steps summarized as follows to solve four sub-problems. Firstly, we construct the behavior-preserving structure, which exactly encodes all the control equivalent supervisors. Secondly, from the attacker's point of view, based on the behavior-preserving structure, we adopt the supervisor synthesis technique to synthesize a structure that encodes all the covert damage strings, which each one works for at least one control equivalent supervisor. The next two steps are carried out from the defender's point of view. Thirdly, based on the extracted covert damage strings, we adopt the supervisor synthesis technique to synthesize a structure that contains all the fortified supervisors. However, it is possible that there are no control commands defined after an observation occurs as a result of synthesis implemented at step 3, and this is inconsistent with the structure of a bipartite supervisor. Thus, fourthly, we perform an iterative pruning procedure on the structure obtained at step 3 to generate a structure that exactly encodes all the fortified supervisors. Finally, we extract one fortified supervisor from the structure computed by this iterative pruning procedure. We remark that the idea of constructing a finite structure to determine the existence of a supervisor satisfying a certain property has already been used in various literature on formal methods, e.g., \cite{arnold2003games}. This idea is also adopted in \cite{YL17range}-\cite{YL15}. However, the fortified supervisors are associated with properties different from those of the supervisors synthesized in \cite{YL17range}-\cite{YL15}, which naturally makes the constructed structures different.

Our previous works \cite{Lin2018}-\cite{TLZS22} on attack synthesis and \cite{LZTWS22,TLZS22net} on networked DES also adopt the supervisor synthesis technique after problem transformation. However, due to the different natures of sub-problems studied in this work, the constructed supervisory control problems are different from those in our previous works. For step 1, none of our previous works has constructed such a behavior-preserving structure. For step 2, we remark that \cite{Lin2018}-\cite{TLS22} target to identify all the covert damage strings for a given supervisor and \cite{TLZS22} targets to identify all the covert damage strings, which each one works for all the observation-consistent supervisors. Thus, the covert damage strings extracted at step 2 of this work are different from those in our previous works. Such a difference naturally makes the approaches of \cite{Lin2018}-\cite{TLZS22} cannot work for our studied problem. Notice that there may be an infinite number of control equivalent supervisors. Thus, adopting the approaches of \cite{Lin2018}-\cite{TLS22} to identify covert damage strings for each control equivalent supervisor is infeasible for our studied problem. Steps 3 and 4 focus on resilient supervisor synthesis against attack instead of attack synthesis or supervisor synthesis against network delays and losses; thus, our constructed supervisory control problems are also different from those in \cite{Lin2018}-\cite{TLZS22net}. We remark that \cite{LZS19b,Zhu2018} propose constraint-based approaches to compute resilient supervisors without using supervisor synthesis techniques. Last but not least, our method involves multiple steps of invocations of the supervisory synthesis technique, distinguishing it from previous algorithms in \cite{Lin2018}-\cite{TLZS22net} that could solve the problem with just a single-step synthesis.

Our contributions are summarized as follows. Firstly, we show the problem of determining the existence of fortified supervisors against covert actuator attacks is decidable. Secondly, we construct a new behavior-preserving structure to encode all the control equivalent supervisors, based on which we design a sound and complete decision process that may serve as a specific fortified supervisor synthesis procedure. We also discuss how to extend the decidability result to the case against the worst-case attack by relying on the behavior-preserving structure constructed in this work and the structure containing all robust supervisors synthesized in \cite{GLM22}.


This paper is organized as follows. We recall the preliminaries in Section \ref{sec:Preliminaries}. In Section \ref{sec:System architecture and Problem Formulation}, we introduce the problem formulation. 
Section \ref{sec:Generation of control equivalent bipartite supervisors} constructs the behavior-preserving structure to encode all the control equivalent supervisors. Section \ref{sec:Synthesis of Fortified Supervisors Against Covert Actuator Attackers}  synthesizes fortified supervisors and shows the decidability result. The extension of the decidability result to the case against the worst-case attack is also discussed in Section \ref{sec:Synthesis of Fortified Supervisors Against Covert Actuator Attackers}. Finally, conclusions are drawn in Section \ref{sec:conclusions}. A running example is given throughout the paper. 


\section{Preliminaries}
\label{sec:Preliminaries}
Let $\mathbb{N}$ be the set of nonnegative integers. Let $[m:n] := \{m,m+1,\cdots,n\}$ ($m \in \mathbb{N}, n \in \mathbb{N}$). 
$\Sigma^{*}$ is the Kleene-closure of a finite alphabet $\Sigma$.
For a string $s$, $|s|$ is defined to be the length of $s$. Given two strings $s, t \in \Sigma^{*}$, we say $s$ is a prefix substring of $t$, written as $s \leq t$, if there exists $u \in \Sigma^{*}$ such that $su = t$. 
The prefix closure of a language $L \subseteq \Sigma^{*}$ is defined as $\overline{L} = \{u \in \Sigma^{*} \mid (\exists v \in L) \, u\leq v\}$. 
$\mathcal{P}_{j}(s)$ represents the prefix of length $j$, specifically, $\mathcal{P}_{0}(\cdot) = \varepsilon$. 
$s[i]$ denotes the $i$-th element in $s$. $s^{\downarrow}$ denotes the last event in $s$.
As usual, $P_{\Sigma'}: \Sigma^{*} \rightarrow (\Sigma')^{*}$ is the natural projection defined as follows: 1) $P_{\Sigma'}(\varepsilon) = \varepsilon$, 2) $(\forall \sigma \in \Sigma) \, P_{\Sigma'}(\sigma) = \sigma$ if $\sigma \in \Sigma'$, otherwise, $P_{\Sigma'}(\sigma) = \varepsilon$, 3) $(\forall s \in \Sigma^{*}, \sigma \in \Sigma) \, P_{\Sigma'}(s\sigma) = P_{\Sigma'}(s)P_{\Sigma'}(\sigma)$.
A finite state automaton $G$ is given by a 5-tuple $(Q, \Sigma, \xi, q_{0}, Q_{m})$, where $Q$ is the state set, $\Sigma$ is the event set, $\xi: Q \times \Sigma \rightarrow Q$ is the (partial) transition function, $q_{0} \in Q$ is the initial state, and $Q_{m}$ is the set of marker states. 
We write $\xi(q, \sigma)!$ to mean that $\xi(q, \sigma)$ is defined. We define $En_{G}(q) = \{\sigma \in \Sigma|\xi(q, \sigma)!\}$.
$\xi$ is extended to the (partial) transition function $\xi: Q \times \Sigma^{*} \rightarrow Q$ by recursively defining $\xi(q,\varepsilon) = q$ and $\xi(q,s\sigma) = \xi(\xi(q, s), \sigma)$, where $q \in Q$, $s \in \Sigma^{*}$ and $\sigma \in \Sigma$, and is extended to the transition function $\xi: 2^{Q} \times \Sigma \rightarrow 2^{Q}$ by defining $\xi(Q', \sigma) = \{\xi(q, \sigma)|q \in Q'\}$, where $Q' \in 2^{Q}$ and $\sigma \in \Sigma$ \cite{WMW10}.
Let $L(G)$ and $L_{m}(G)$ denote the closed-behavior and the marked behavior, respectively. $G$ is said to be marker-reachable if $L_m(G) \neq \varnothing$ \cite{WMW10}. When $Q_{m} = Q$, we shall also write $G = (Q, \Sigma, \xi, q_{0})$ for simplicity. $Ac(G)$ stands for the automaton by deleting those states (and the associated transitions) that are not reachable from the initial state in $G$ \cite{CL99}.
The ``unobservable reach'' of the state $q \in Q$ under the subset of events $\Sigma' \subseteq \Sigma$ is given by $UR_{G, \Sigma - \Sigma'}(q) := \{q' \in Q|[\exists s \in (\Sigma - \Sigma')^{*}] \, q' = \xi(q,s)\}$.
We define $\mathscr{P}_{\Sigma'}(G)$ to be the finite state automaton $(2^{Q} - \{\varnothing\}, \Sigma, \delta, UR_{G, \Sigma - \Sigma'}(q_{0}))$ over $\Sigma$, where the (partial) transition function $\delta: (2^{Q} - \{\varnothing\}) \times \Sigma \rightarrow (2^{Q} - \{\varnothing\})$ is defined as follows:
\begin{enumerate}[1)]
\setlength{\itemsep}{3pt}
\setlength{\parsep}{0pt}
\setlength{\parskip}{0pt}
    \item For any $\varnothing \neq Q' \subseteq Q$ and any $\sigma \in \Sigma'$, if $\xi(Q', \sigma) \neq \varnothing$, then $\delta(Q', \sigma) = UR_{G, \Sigma - \Sigma'}(\xi(Q', \sigma))$, where $UR_{G, \Sigma - \Sigma'}(Q'') = \bigcup_{q \in Q''}UR_{G, \Sigma - \Sigma'}(q)$
    for any $\varnothing \neq Q'' \subseteq Q$;
    \item For any $\varnothing \neq Q' \subseteq Q$ and any $\sigma \in \Sigma - \Sigma'$, if there exists $q \in Q'$ such that $\xi(q, \sigma)!$, then $\delta(Q', \sigma) = Q'$. 
\end{enumerate}


For $G = (Q, \Sigma, \xi, q_{0}, Q_{m})$, after removing the states in $Q' \subseteq Q$ and the transitions associated with the states in $Q' \subseteq Q$, the generated automaton is denoted as $G^{|Q-Q'}$.
For any two finite state automata $G_{1}$ and $G_{2}$, their parallel composition \cite{CL99} is denoted as $G_{1}||G_{2}$.

Following \cite{WMW10,CL99}, for a plant modeled as a finite state automaton $G = (Q, \Sigma, \xi, q_{0}, Q_{m})$, its event set $\Sigma$ is partitioned into $\Sigma = \Sigma_{c} \dot{\cup} \Sigma_{uc} = \Sigma_{o} \dot{\cup} \Sigma_{uo}$, where $\Sigma_{c}$ ($\Sigma_{o}$) and $\Sigma_{uc}$ ($\Sigma_{uo}$) are defined as the sets of controllable (observable) and uncontrollable (unobservable) events, respectively. A control constraint over $\Sigma$ is a tuple $(\Sigma_{c}, \Sigma_{o})$, which specifies the control and observation capability of a supervisor on the plant. A supervisory control map of $G$ over the control constraint $(\Sigma_{c}, \Sigma_{o})$ is defined as $V: L(G) \rightarrow \Gamma$, where $\Gamma := \{\gamma \subseteq \Sigma|\Sigma_{uc} \subseteq \gamma\}$, such that $\forall s, t \in L(G): P_{\Sigma_{o}}(s) = P_{\Sigma_{o}}(t) \Rightarrow V(s) = V(t)$. 
Let $V/G$ denote the closed-loop system under supervision of $V$, which is defined as follows: 1) $\epsilon \in L(V/G)$, 2) for any $s \in L(V/G)$ and $\sigma \in \Sigma$, $s\sigma \in L(V/G) \Leftrightarrow s\sigma \in L(G) \wedge \sigma \in V(s)$, and 3) $L_{m}(V/G) := L_{m}(G) \cap L(V/G)$.
The control map $V$ is finitely representable if $V$ can be described by a finite state automaton, say $S = (Q_{s}, \Sigma, \xi_{s}, q_{s}^{init})$, such that 1) $L(S||G) = L(V/G)$ and $L_{m}(S||G) = L_{m}(V/G)$, 2) for any $s \in L(S)$, $En_{S}(\xi_{s}(q_{s}^{init},s)) = V(s)$, and 3) for any $s,t \in L(S)$, $P_{\Sigma_{o}}(s) = P_{\Sigma_{o}}(t) \Rightarrow \xi_{s}(q_{s}^{init},s) = \xi_{s}(q_{s}^{init},t)$. The \emph{Basic Supervisory Control and Observation Problem (BSCOP)} is as follows.

\textbf{Definition II.1 (BSCOP) \cite{CL99}:} Given plant $G$ and legal language $L_{a} = \overline{L_{a}}$, find a supervisor $S$ over the control constraint $(\Sigma_{c}, \Sigma_{o})$ such that 1) $L(S||G) \subseteq L_{a}$, and 2) for any other supervisor $S'$ such that $L(S'||G) \subseteq L_{a}$, $L(S'||G) \subseteq L(S||G)$. 

When $\Sigma_{c} \subseteq \Sigma_{o}$, the supremal\footnote{This statement follows the standard Ramadge-Wonham control terminology \cite{WMW10,CL99}, where the supremal supervisor can also be empty, whose state
space is empty and consequently whose generated and marked languages are empty. We refer readers to \cite{WMW10,CL99} for more details about the supremality.} solution always exists for BSCOP \cite{CL99}, although it may be empty.
Table \ref{tab:notations} summarizes the notations of main components and symbols used in this work.
\begin{table}[htbp]  
  \centering  
  \caption{NOTATIONS}  
  \label{tab:notations}  
  \begin{tabular}{ll}  
    \hline  
    \hline\\ [-0.3cm] 
    Notation & Meaning\\  
    \hline 
    $G$    &  Plant\\  
    \hline
    $S$    & Supervisor\\  
    \hline
    $CE$  &   Command execution\\  
    \hline
    $CE^{A}$  &   Command execution under attack\\  
    \hline
    $BT(S)$    & Bipartization of $S$ (Bipartite supervisor)\\
    \hline
    $BT(S)^{A}$    & \tabincell{l}{Bipartization of $S$ under attack (Bipartite \\ supervisor under attack)}\\
    \hline
    $\mathcal{A}$          &  Actuator attacker\\  
    \hline
    $BPS(S)$    & Bipartite behavior-preserving structure \\
    \hline
    $BPNS(S)$    & \tabincell{l}{Bipartite behavior-preserving \\command-nondeterministic supervisor}\\
    \hline
    $BPNS^{A}(S)$    & \tabincell{l}{Bipartite behavior-preserving \\ command-nondeterministic supervisor \\ under attack}\\
    \hline
    $\hat{\mathcal{A}}$    & \tabincell{l}{The structure that encodes all the covert \\ damage strings}\\
    \hline
    $FNS(S)$    & \tabincell{l}{Fortified command-nondeterministic \\ supervisor} \\
    \hline
    $FS(S)$    & \tabincell{l}{Fortified supervisor extracted from \\ $FNS(S)$} \\
    \hline
    $\Sigma_{o,a}$  &   \tabincell{l}{The set of observable (plant) events for \\ the attacker}\\
    \hline
    $\Sigma_{c,a}$  &   \tabincell{l}{The set of actuator attackable events for \\ the attacker}\\
    \hline  
    \hline  
  \end{tabular}  
\end{table}



\section{Component Models and Problem Formulation}
\label{sec:System architecture and Problem Formulation}

In this section, we firstly present the system architecture under actuator attack and describe the problem to be solved in plain language. Then, we formally define the component models, based on which we formulate the studied problem.


\subsection{Component models}
\label{subsec:Component models}

\begin{figure}[htp]
\begin{center}
\includegraphics[height=1.8cm]{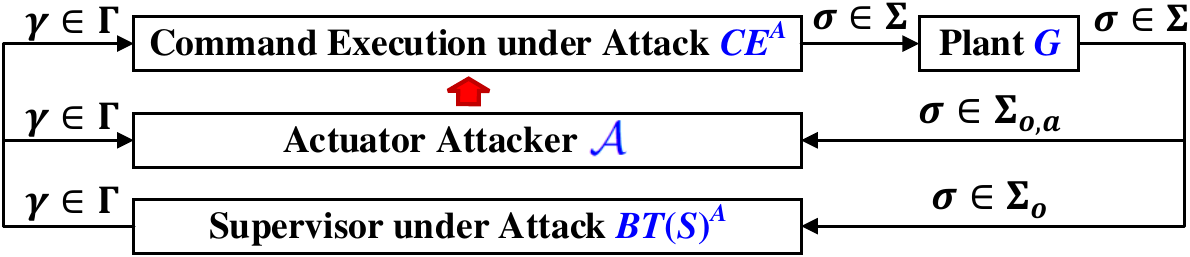}   
\caption{Supervisory control architecture under actuator attack}
\label{fig:architecture}
\end{center}        
\end{figure}

The supervisory control architecture under actuator attack is shown in Fig. \ref{fig:architecture}, consisting of the following components:
\begin{itemize}
\setlength{\itemsep}{3pt}
\setlength{\parsep}{0pt}
\setlength{\parskip}{0pt}
    \item Plant: The plant is a controlled system. There are some damage states in the plant, which should be avoided under the control of a supervisor.
    \item Supervisor under attack: The supervisor issues one control command in $\Gamma = \{\gamma \subseteq \Sigma|\Sigma_{uc} \subseteq \gamma\}$ whenever an event in $\Sigma_{o}$ is received or when the system initiates. Once the supervisor sees any unexpected observable sequence, it asserts that the information inconsistency happens and an attack is detected \cite{Su2018,Su20}.
    \item Command execution automaton under attack: This component executes an event when a control command is received, which is compromised by the actuator attack, i.e., it describes the attacked phase from control command reception to event execution at the plant.
    \item Actuator attacker: The actuator attacker can modify each control command by enabling or disabling the controllable events in $\Sigma_{c,a} \subseteq \Sigma_{c}$. The set of observable events in $\Sigma$ for the actuator attacker is denoted by $\Sigma_{o,a} \subseteq \Sigma$. The control commands in $\Gamma$ are observable to the actuator attacker. We assume that $\Sigma_{c,a} \subseteq \Sigma_{o,a}$. In this work, the actuator attacker is nondeterministic, i.e., there might exist more than one attack choice over each observation, and covert, i.e., the attack operations cannot initiated by the attacker will always prevent information inconsistency from occurring at the supervisor side. The attacker's goal is to induce the plant to reach damage states.
\end{itemize}
Before we present the formal definition of each component and the studied problem, we first describe the problem in plain language. Consider a supervisor $S$ for which at least a covert actuator attacker could induce the plant to reach damage states. The problem is whether there exists a new supervisor $S'$ such that 1) the closed-loop system behavior under the control of $S$ can be preserved under the control of $S'$, and 2) there does not exist any covert actuator attacker that could induce the plant to reach damage states for $S'$.

\textbf{Remark III.1:} The assumptions $\Sigma_{o,a} \subseteq \Sigma_{o}$ and $\Sigma_{c} \subseteq \Sigma_{o}$ in \cite{Zhu2018} are relaxed in this work. Thus, we consider a more general setup.
The assumption $\Sigma_{c,a} \subseteq \Sigma_{o,a}$ enables us to synthesize the supremal structure in Section \ref{subsec:Damage strings encoding} to encode all the covert damage strings, which would be explained later. Without this assumption, there might exist an infinite number of maximal permissive structures to encode those damage strings, resulting in that the later control command pruning procedure in Section \ref{subsec:Illegal control commands pruning} cannot terminate within finite steps and the decidability result becomes tricky to be established.

Next, following \cite{LZS19}, we explain how to model these components as finite state automata.

\subsubsection{Plant}
\label{subsubsec:Plant}

The plant is modeled as a finite state automaton $G = (Q, \Sigma, \xi, q^{init}, Q_{d})$, where $Q_{d}$ is the set of damage states. In the following text, for the consistency in notations, we use $q^{init}$ to denote the initial state of the plant.

\subsubsection{Supervisor}
\label{subsubsec:Supervisor}

The supervisor is modeled as a finite state automaton $S = (Q_{s}, \Sigma, \xi_{s}, q_{s}^{init})$ satisfying the controllability and observability constraints \cite{B1993}: 1) (controllability) $(\forall q \in Q_{s})(\forall \sigma \in \Sigma_{uc})\xi_{s}(q, \sigma)!$, and 2) (observability) $(\forall q \in Q_{s})(\forall \sigma \in \Sigma_{uo})\xi_{s}(q, \sigma)! \Rightarrow \xi_{s}(q, \sigma) = q$.
The control command issued by $S$ at state $q \in Q_{s}$ is defined to be $\Gamma(q) := En_{S}(q) = \{\sigma \in \Sigma|\xi_{s}(q,\sigma)!\} \subseteq \Gamma$. 

Next, we perform a bipartization transformation on $S$ such that the observation reception phase and control command sending phase become explicit while the control function remains the same as $S$. We call the transformed structure a bipartite supervisor \cite{LZS19}, denoted by $BT(S) = (Q_{bs}, \Sigma_{bs}, \xi_{bs}, q_{bs}^{init})$, where:
\begin{itemize}
\setlength{\itemsep}{3pt}
\setlength{\parsep}{0pt}
\setlength{\parskip}{0pt}
    \item $Q_{bs} = Q_{s} \cup Q_{s}^{com}$, where $Q_{s}^{com} = \{q^{com} \mid q \in Q_s\}$.
    \item $q_{bs}^{init} = (q_{s}^{init})^{com}$
    \item $\Sigma_{bs} = \Sigma \cup \Gamma$
    \item $\xi_{bs}$ is defined as:
    \begin{enumerate}[1.]
    \setlength{\itemsep}{3pt}
    \setlength{\parsep}{0pt}
    \setlength{\parskip}{0pt}
        \item $(\forall q^{com} \in Q_{s}^{com}) \, \xi_{bs}(q^{com}, \Gamma(q)) = q$.
        \item $(\forall q \in Q_{s})(\forall \sigma \in \Sigma_{uo}) \, \xi_{s}(q, \sigma)! \Rightarrow \xi_{bs}(q, \sigma) = q$.
        \item $(\forall q \in Q_{s})(\forall \sigma \in \Sigma_{o}) \, \xi_{s}(q, \sigma)! \Rightarrow \xi_{bs}(q, \sigma) = (\xi_{s}(q, \sigma))^{com}$. 
    \end{enumerate}
\end{itemize}
In the state set $Q_{bs}$, any $q \in Q_{s}$ is a reaction state ready to observe any event in $\Gamma(q)$, and any $q^{com} \in Q_{s}^{com}$ is a control state corresponding to $q$, which is ready to issue the control command $\Gamma(q)$. Thus, the initial state is $(q_{s}^{init})^{com}$. For $\xi_{bs}$, 1) at any control state $q^{com}$, a control command $\Gamma(q)$ should be issued, which leads to a reaction state $q$ (Case 1), and 2) at any reaction state $q$, any unobservable event, if defined in $S$, is a self-loop (Case 2), and any observable event, if defined in $S$, would lead to a control state $(\xi_{s}(q, \sigma))^{com}$ (Case 3).

For the bipartite supervisor $BT(S)$, we now model the effect of actuator attacks on $BT(S)$ by constructing the bipartite supervisor under attack, denoted by $BT(S)^{A} = (Q_{bs}^{a}, \Sigma_{bs}^{a}, \xi_{bs}^{a}, q_{bs}^{a,init})$, where: 
\begin{itemize}
\setlength{\itemsep}{3pt}
\setlength{\parsep}{0pt}
\setlength{\parskip}{0pt}
    \item $Q_{bs}^{a} = Q_{bs} \cup \{q^{detect}\}$
    \item $q_{bs}^{a,init} = q_{bs}^{init}$
    \item $\Sigma_{bs}^{a} = \Sigma \cup \Gamma$
    \item $\xi_{bs}^{a}$ is defined as:
    \begin{enumerate}[1.]
    \setlength{\itemsep}{3pt}
    \setlength{\parsep}{0pt}
    \setlength{\parskip}{0pt}
        \item $(\forall q, q' \in Q_{bs}^{a})(\forall \sigma \in \Sigma \cup \Gamma) \xi_{bs}(q, \sigma) = q' \Rightarrow \xi_{bs}^{a}(q, \sigma) = q'$
        \item $(\forall q \in Q_{s})(\forall \sigma \in \Sigma_{c,a} \cap \Sigma_{uo}) \neg\xi_{bs}(q, \sigma)! \Rightarrow \xi_{bs}^{a}(q, \sigma) = q$ 
        \item $(\forall q \in Q_{s})(\forall \sigma \in \Sigma_{o}) \neg\xi_{bs}(q, \sigma)! \Rightarrow \xi_{bs}^{a}(q, \sigma) = q^{detect}$ 
    \end{enumerate}
\end{itemize}
In the state set $Q_{bs}^{a}$, a new state $q^{detect}$ is added, denoting the situation where the actuator attack is detected. For $\xi_{bs}^{a}$, Case 1 retains all the transitions defined in $BT(S)$. In Case 2, for any reaction state $q \in Q_{s}$, the transitions labelled by unobservable and attackable events in $\Sigma_{c,a} \cap \Sigma_{uo}$, which are not defined at the state $q$ in $BT(S)$, are added. In Case 3, for any reaction state $q \in Q_{s}$, the transitions labelled by observable events, which are not defined at the state $q$ in $BT(S)$, would lead to the state $q^{detect}$, with the interpretation that the supervisor has received some observation that should not have occurred based on the supervisor structure, i.e., the actuator attacker is detected.

\subsubsection{Command execution automaton}
\label{subsubsec:command execution automaton}

To explicitly encode the phase from receiving a control command in $\Gamma$ to executing an event in $\Sigma$ at the plant, we construct the command execution automaton \cite{LZS19} $CE = (Q_{ce}, \Sigma_{ce}, \xi_{ce}, q_{ce}^{init})$, where 
\begin{itemize}
\setlength{\itemsep}{3pt}
\setlength{\parsep}{0pt}
\setlength{\parskip}{0pt}
    \item $Q_{ce} = \{q^{\gamma}|\gamma \in \Gamma\} \cup \{q_{ce}^{init}\}$
    \item $\Sigma_{ce} = \Gamma \cup \Sigma$
    \item $\xi_{ce}$ is defined as:
    \begin{enumerate}[1.]
    \setlength{\itemsep}{3pt}
    \setlength{\parsep}{0pt}
    \setlength{\parskip}{0pt}
        \item $(\forall \gamma \in \Gamma) \xi_{ce}(q_{ce}^{init}, \gamma) = q^{\gamma}$
        \item $(\forall \gamma \in \Gamma)(\forall \sigma \in \gamma \cap \Sigma_{o}) \xi_{ce}(q^{\gamma}, \sigma) = q_{ce}^{init}$
        \item $(\forall \gamma \in \Gamma)(\forall \sigma \in \gamma \cap \Sigma_{uo}) \xi_{ce}(q^{\gamma}, \sigma) = q^{\gamma}$
    \end{enumerate}
\end{itemize}
The above command execution automaton $CE$ is an unattacked model, i.e., the event execution always follows the unmodified control command based on $CE$. We need to encode the actuator attack effects into $CE$ to build the model for the command execution process under attack, denoted by $CE^{A} = (Q_{ce}^{a}, \Sigma_{ce}^{a}, \xi_{ce}^{a}, q_{ce}^{a,init})$, where 
\begin{itemize}
\setlength{\itemsep}{3pt}
\setlength{\parsep}{0pt}
\setlength{\parskip}{0pt}
    \item $Q_{ce}^{a} = Q_{ce}$
    \item $q_{ce}^{a,init} = q_{ce}^{init}$
    \item $\Sigma_{ce}^{a} = \Gamma \cup \Sigma$
    \item $\xi_{ce}^{a}$ is defined as:
    \begin{enumerate}[1.]
    \setlength{\itemsep}{3pt}
    \setlength{\parsep}{0pt}
    \setlength{\parskip}{0pt}
        \item $(\forall q, q' \in Q_{ce}^{a})(\forall \sigma \in \Sigma \cup \Gamma) \xi_{ce}(q, \sigma) = q' \Rightarrow \xi_{ce}^{a}(q, \sigma) = q'$ 
        \item $(\forall \gamma \in \Gamma)(\forall \sigma \in \Sigma_{c,a} \cap \Sigma_{o}) \neg\xi_{ce}(q^{\gamma}, \sigma)! \Rightarrow \xi_{ce}^{a}(q^{\gamma}, \sigma) = q_{ce}^{a,init}$
        \item $(\forall \gamma \in \Gamma)(\forall \sigma \in \Sigma_{c,a} \cap \Sigma_{uo}) \neg\xi_{ce}(q^{\gamma}, \sigma)! \Rightarrow \xi_{ce}^{a}(q^{\gamma}, \sigma) = q^{\gamma}$
    \end{enumerate}
\end{itemize}
For $\xi_{ce}^{a}$, Case 1 retains all the transitions defined in $CE$. In Case 2 and Case 3, the enablement attacks are encoded. For any state $q^{\gamma}$, the transitions labelled by attackable events, which are not defined at the state $q^{\gamma}$ in $CE$, are added, where the observable events would lead to the initial state $q_{ce}^{init}$ (Case 2), and the unobservable events would lead to self-loop transitions (Case 3). Here we remark that $CE^{A}$ only needs to introduce actuator enablement attacks, and the actuator disablement would be automatically taken care of by the synthesis procedure in the latter \textbf{Procedure 1} as the attackable event set $\Sigma_{c,a}$ is controllable by the actuator attacker, i.e., the actuator attack could always disable these events.

\subsubsection{Actuator attacker}
\label{subsubsec:Actuator attacker}

To formulate the closed-loop system under attack, which is later used for several formal definitions in Section \ref{subsec:Problem formulation}, we also model the actuator attacker as a finite state automaton $\mathcal{A} = (Q_{a}, \Sigma_{a}, \xi_{a}, q_{a}^{init})$, where $\Sigma_{a} = \Sigma \cup \Gamma$. There are two conditions that need to be satisfied:
\begin{itemize}
\setlength{\itemsep}{3pt}
\setlength{\parsep}{0pt}
\setlength{\parskip}{0pt}
    \item ($\mathcal{A}$-controllability) For any state $q \in Q_{a}$ and any event $\sigma \in \Sigma_{a,uc} := \Sigma_{a} - \Sigma_{c,a}$, $\xi_{a}(q, \sigma)!$ 
    \item ($\mathcal{A}$-observability) For any state $q \in Q_{a}$ and any event $\sigma \in \Sigma_{a,uo} := \Sigma_{a} - (\Sigma_{o,a} \cup \Gamma)$, if $\xi_{a}(q, \sigma)$!, then $\xi_{a}(q, \sigma) = q$.
\end{itemize}
$\mathcal{A}$-controllability states that the attacker can only disable events in $\Sigma_{c,a}$. $\mathcal{A}$-observability states that the attacker can only make a state change after observing an event in $\Sigma_{o,a} \cup \Gamma$. In the following text, we shall refer to $(\Sigma_{o,a}, \Sigma_{c,a})$ as the attack constraint, and $\mathscr{C}_{ac} = (\Sigma_{c,a}, \Sigma_{o,a} \cup \Gamma)$ as the attacker's control constraint. 
The model of $\mathcal{A}$ is nondeterministic in terms of making attack decisions as it allows multiple attack choices upon each observation.

\subsection{Problem formulation}
\label{subsec:Problem formulation}

As shown in Fig. \ref{fig:architecture}, the closed-loop system under attack can be modeled as the parallel composition of the above-constructed component models, denoted by $CLS^{A} = G||CE^{A}||BT(S)^{A}||\mathcal{A} = (Q_{cls}^{a}, \Sigma_{cls}^{a}, \xi_{cls}^{a}, q_{cls}^{a,init}, Q_{cls,m}^{a})$.

\textbf{Definition III.1 (Covertness):} Given $G$, $BT(S)^{A}$ and $CE^{A}$, an actuator attacker $\mathcal{A}$ is said to be covert against $S$ w.r.t. the  attack constraint $(\Sigma_{o,a}, \Sigma_{c,a})$ if any state in $\{(q_{g},q_{ce}^{a},q_{bs}^{a}, q_{a}) \in Q_{cls}^{a}| q_{bs}^{a} = q^{detect}\}$ is not reachable in $CLS^{A}$.

\textbf{Remark III.2:} The notion of covertness depends on the monitoring mechanism used. The above covertness definition (following \cite{Su2018,Su20}) means that the supervisor relies on its own structure to monitor the information inconsistency. Notice that we could also rely on the observer of normal behavior without attacks to monitor attacks, similar to \cite{LS20J}. However, compared to the observer-based monitoring mechanism, the covertness definition adopted in our work allows more attacks to become covert. Consequently, our synthesized fortified supervisor exhibits resilience against a wider range of attacks.

\textbf{Remark III.3:} When a smart attacker is deployed in the system, it remains hidden in the system as long as possible so that it can obtain as much information as possible and attempt as much damage infliction as possible through repetitive system runs. From such a point of view, considering the resilient control against smart attackers that could remain covert at all times, captured by \textbf{Definition III.1}, is practically essential. 

\textbf{Definition III.2 (Damage-reachable):} Given $G$, $BT(S)^{A}$ and $CE^{A}$, an actuator attacker $\mathcal{A}$ is said to be damage-reachable against $S$ w.r.t. the  attack constraint $(\Sigma_{o,a}, \Sigma_{c,a})$ if $L_{m}(CLS^{A}) \neq \varnothing$.

\textbf{Definition III.3 (Resilience):} Given $G$, a supervisor $S$ is said to be resilient if there does not exist any covert and damage-reachable actuator attacker $\mathcal{A}$ against $S$ w.r.t. the attack constraint $(\Sigma_{o,a}, \Sigma_{c,a})$. 

In this work, we assume that the original supervisor, denoted as $S$, is non-resilient for the plant $G$. 

\textbf{Definition III.4 (Control equivalence):} Given $G$ and $S$, a supervisor $S'$ (bipartite supervisor $BT(S')$, respectively) is said to be control equivalent to $S$ ($BT(S)$, respectively) if $L(G||S) = L(G||S')$ ($P_{\Sigma}(L(G||CE||BT(S))) = P_{\Sigma}(L(G||CE||BT(S')))$, respectively).

\textbf{Definition III.5 (Fortification):} Given $G$ and a non-resilient $S$, a supervisor $S'$ is said to be a fortified supervisor for $S$ if $S'$ is resilient and control equivalent to $S$.

\textbf{Definition III.6 (Covert damage string):} Given $G$, a non-resilient $S$, and a covert damage-reachable actuator attacker $\mathcal{A}$ against $S$, any $s \in L_{m}(G||CE^{A}||BT(S)^{A}||\mathcal{A})$ is a covert damage string that works for $S$.

Now we are ready to formulate the studied problem.

\textbf{Problem 1:} Given $G$, a non-resilient $S$, and the attack constraint $(\Sigma_{o,a}, \Sigma_{c,a})$, determine whether there exists a fortified supervisor for $S$.

\begin{figure}[htbp]
\centering

\subfigure[]{
\begin{minipage}[t]{0.5\linewidth}
\centering
\includegraphics[height=0.64in]{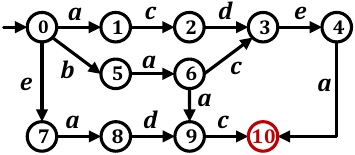}
\end{minipage}%
}%
\subfigure[]{
\begin{minipage}[t]{0.5\linewidth}
\centering
\includegraphics[height=0.3in]{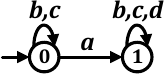}
\end{minipage}%
}%

\subfigure[]{
\begin{minipage}[t]{0.5\linewidth}
\centering
\includegraphics[height=0.65in]{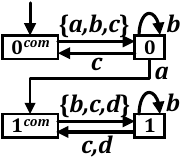}
\end{minipage}
}%
\subfigure[]{
\begin{minipage}[t]{0.5\linewidth}
\centering
\includegraphics[height=0.65in]{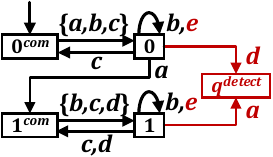}
\end{minipage}
}%

\centering
\caption{(a) $G$. (b) $S$. (c) $BT(S)$. (d) $BT(S)^{A}$.}
\label{fig:G_S_BTS_BTSA}
\end{figure}

\begin{figure}[htbp]
\centering
\subfigure[]{
\begin{minipage}[t]{1\linewidth}
\centering
\includegraphics[height=1.3in]{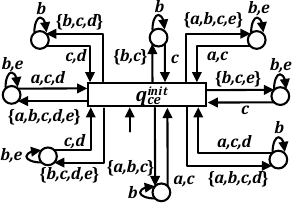}
\end{minipage}
}
\subfigure[]{
\begin{minipage}[t]{1\linewidth}
\centering
\includegraphics[height=1.3in]{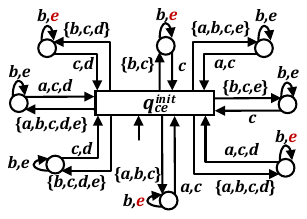}
\end{minipage}
}

\centering
\caption{(a) $CE$. (b) $CE^{A}$.}
\label{fig:CE_CEA}
\end{figure}

\textbf{Example III.1} Consider the plant $G$ and supervisor $S$ shown in Fig. \ref{fig:G_S_BTS_BTSA}. $\Sigma = \{a,b,c,d,e\}$. $\Sigma_{o} = \{a,c,d\}$. $\Sigma_{uo} = \{b,e\}$. $\Sigma_{c} = \{a,d,e\}$. $\Sigma_{uc} = \{b,c\}$. $\Sigma_{o,a} = \{b,c,d,e\}$. $\Sigma_{c,a} = \{e\}$. The damage state is state 10, i.e., $Q_{d} = \{10\}$. We have $L(G||S) = \overline{\{acd,bac\}}$.
Fig. \ref{fig:G_S_BTS_BTSA} and Fig. \ref{fig:CE_CEA} illustrate $BT(S)$, $BT(S)^{A}$, $CE$, and $CE^{A}$, where the differences between $BT(S)$ and $BT(S)^{A}$, and $CE$ and $CE^{A}$ are marked in red. It can be checked that $S$ is non-resilient as covert and damage-reachable actuator attackers exist, e.g., enabling the event $e$ after the initial control command $\{a,b,c\}$. 





\section{Main idea of the solution methodology}
\label{sec:Main idea of the solution methodology}

\begin{figure}[htp]
\begin{center}
\includegraphics[height=5.9cm]{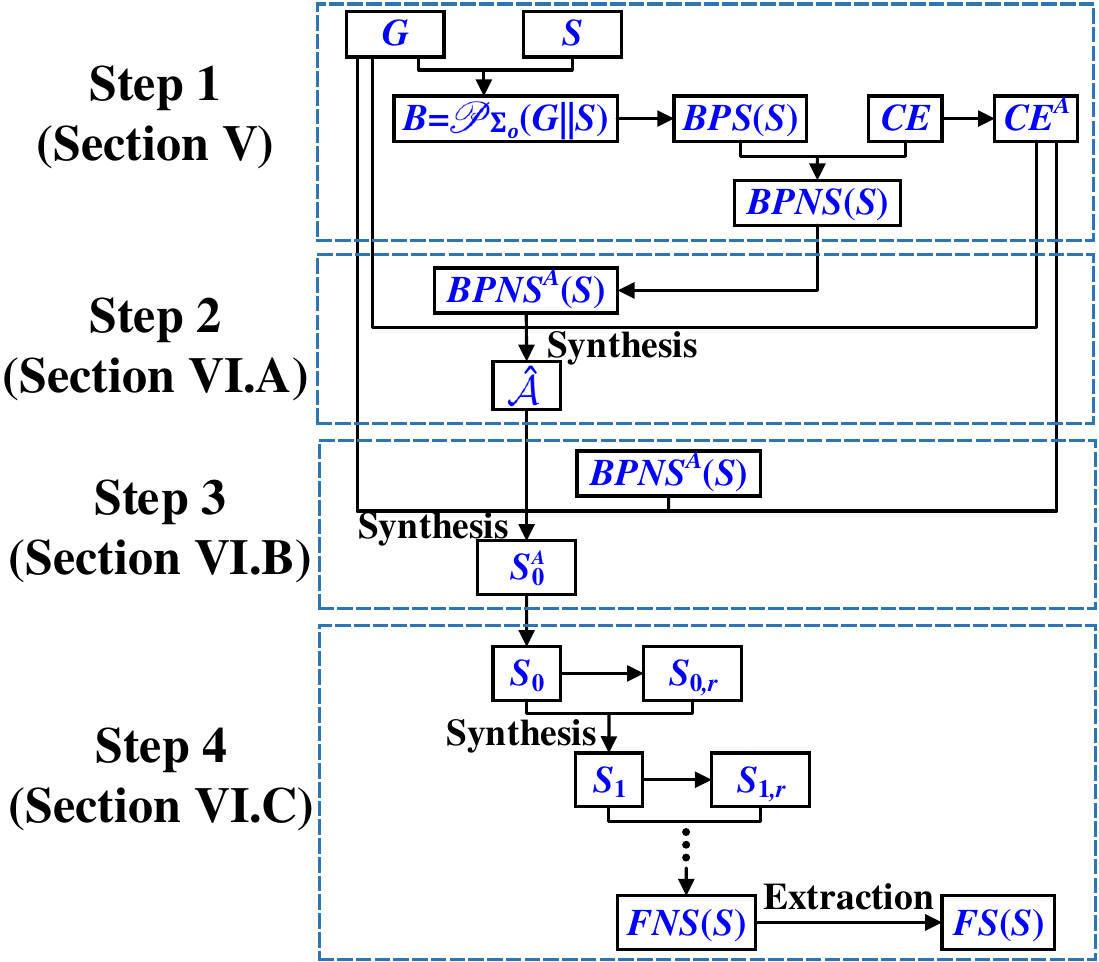}   
\caption{The procedure of the proposed solution methodology}
\label{fig:Solution_Methodology}
\end{center}        
\end{figure}

Before proceeding further, we first present the high-level idea, as illustrated in Fig. \ref{fig:Solution_Methodology}.
In Step 1 (Section \ref{sec:Generation of control equivalent bipartite supervisors}), based on $G$, $S$, and $CE$, we construct the behavior-preserving structure $BPNS(S)$ to exactly encode all the control equivalent bipartite supervisors. In Step 2 (Section \ref{subsec:Damage strings encoding}), based on $G$, $CE^{A}$ and $BPNS^{A}(S)$, which is the version of $BPNS(S)$ under attack, we synthesize $\hat{\mathcal{A}}$ to encode all the covert damage strings. In Step 3 (Section \ref{subsec:Illegal control commands pruning}), based on $G$, $CE^{A}$, $BPNS^{A}(S)$ and $\hat{\mathcal{A}}$, we prune from $BPNS^{A}(S)$ those illegal control commands leading to damage infliction to generate $S_{0}^{A}$. In Step 4 (Section \ref{subsec:Generation of Fortified Supervisors}), we iteratively prune from $S_{0}$ (the non-attacked version of $S_{0}^{A}$) those states where there is no control command defined upon each observation, and generate $FNS(S)$, which exactly encodes all the fortified supervisors. Finally, we extract one supervisor $FS(S)$ from $FNS(S)$, which is the solution. 

\section{Behavior-preserving structure construction}
\label{sec:Generation of control equivalent bipartite supervisors}

In this section, we introduce how to construct the behavior-preserving structure to exactly encode all the control equivalent bipartite supervisors. 

\vspace{-0.2cm}

\subsection{Equivalent behavior computation}
\label{subsec:Equivalent behavior computation}

Firstly, we compute $G||S$, which is the closed-loop system under $S$ in the absence of attacks. 
Recall that for any supervisor $S'$, we have $L(BT(S')) \subseteq \overline{(\Gamma\Sigma_{uo}^{*}\Sigma_{o})^{*}}$ and any unobservable event defined in $BT(S')$ is a self-loop transition. Since we need to construct a structure to encode all the control equivalent bipartite supervisors, we compute a subset construction $B = \mathscr{P}_{\Sigma_{o}}(G||S) = (Q_{b}, \Sigma_{b} = \Sigma, \xi_{b}, q_{b}^{init})$, where $|Q_{b}| \leq 2^{|Q| \times |Q_{s}|}$. By construction, $B$ is built upon the observer \cite{CL99} of $G||S$ by adding self-loops labelled by the unobservable events that could occur at each state of the observer of $G||S$. 
It can be checked that, for any $(q_{g}, q_{s}), (q_{g}', q_{s}') \in q \in Q_{b}$, we have $q_{s} = q_{s}'$ since all the unobservable events in $\Sigma_{uo}$ are self-loops in $S$. Thus, for any $q_{1}, q_{2} \in Q_{b}$ and any $\sigma \in \Sigma_{o}$ such that $\xi_{b}(q_{1}, \sigma) = q_{2}$, where $q_{1,s}$ ($q_{2,s}$, respectively) is the supervisor state in the state $q_{1}$ ($q_{2}$, respectively), we have 1) the state of the supervisor $S$ transits from $q_{1,s}$ to $q_{2,s}$ upon the observation of $\sigma$, and 2) $En_{B}(q_{2})$ contains all the events that could happen at $G$ when $S$ issues the corresponding control command at the state $q_{2,s}$. Consequently, based on the structure $B$, we could find all the feasible control commands w.r.t. each observation, which would be realized later in Section \ref{subsec:Feasible control commands completion}.

\begin{figure}[htbp]
\begin{center}
\includegraphics[height=0.8cm]{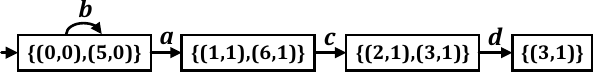}   
\caption{The computed automaton $B$}
\label{fig:B}
\end{center}
\end{figure}

\textbf{Example V.1} Given $G$ and $S$ shown in Fig. \ref{fig:G_S_BTS_BTSA}, the automaton $B$ is shown in Fig. \ref{fig:B}.

\vspace{-0.5cm}

\subsection{Feasible control commands completion}
\label{subsec:Feasible control commands completion}

Based on $B = \mathscr{P}_{\Sigma_{o}}(G||S)$, we generate a bipartite structure similar to $BT(S)$, where upon each observation in $B$, we add all the feasible control commands, under which the closed-behavior of the closed-loop system $L(G||S)$ is preserved. We call such a structure the bipartite behavior-preserving structure, denoted by $BPS(S) = (Q_{bps}, \Sigma_{bps}, \xi_{bps}, q_{bps}^{init})$, where 
\begin{itemize}
\setlength{\itemsep}{3pt}
\setlength{\parsep}{0pt}
\setlength{\parskip}{0pt}
    \item $Q_{bps} = Q_{b} \cup Q_{b}^{com} \cup \{q^{dump}\}$, where $Q_{b}^{com} = \{q^{com}|q \in Q_{b}\}$
    \item $q_{bps}^{init} = (q_{b}^{init})^{com}$
    \item $\Sigma_{bps} = \Sigma \cup \Gamma$
    \item $\xi_{bps}$ is defined as:
    \begin{enumerate}[1.]
    \setlength{\itemsep}{3pt}
    \setlength{\parsep}{0pt}
    \setlength{\parskip}{0pt}
        \item $(\forall q \in Q_{b})(\forall \gamma \in \Gamma)\mathcal{C}_{1} \wedge \mathcal{C}_{2} \Rightarrow \xi_{bps}(q^{com}, \gamma) = q$, where
        \begin{enumerate}[i.]
        \setlength{\itemsep}{3pt}
        \setlength{\parsep}{0pt}
        \setlength{\parskip}{0pt}
            \item $\mathcal{C}_{1} := En_{B}(q) \subseteq \gamma$
            \item $\mathcal{C}_{2} := (\forall (q_{g},q_{s}) \in q)En_{G}(q_{g}) \cap \gamma \subseteq En_{B}(q)$
        \end{enumerate} 
        \item $(\forall q \in Q_{b})(\forall \sigma \in \Sigma_{uo})\xi_{b}(q, \sigma)! \Rightarrow \xi_{bps}(q, \sigma) = q$
        \item $(\forall q \in Q_{b})(\forall \sigma \in \Sigma_{o})\xi_{b}(q, \sigma)! \Rightarrow \xi_{bps}(q, \sigma) = (\xi_{b}(q, \sigma))^{com}$
        \item $(\forall q \in Q_{b})(\forall \sigma \in \Sigma_{uo})\neg\xi_{b}(q, \sigma)! \Rightarrow \xi_{bps}(q, \sigma) = q$
        \item $(\forall q \in Q_{b})(\forall \sigma \in \Sigma_{o})\neg\xi_{b}(q, \sigma)! \Rightarrow \xi_{bps}(q, \sigma) = q^{dump}$
        \item $(\forall \sigma \in \Sigma \cup \Gamma)\xi_{bps}(q^{dump}, \sigma) = q^{dump}$
    \end{enumerate}
\end{itemize}
In the state set $Q_{bps}$, any state $q^{com} \in Q_{b}^{com}$ is a control state corresponding to state $q$, which is ready to issue the control command, and any state $q$ in $Q_{b}$ is a reaction state, which is ready to receive an observation. After a control command is issued at a control state $q^{com}$, $BPS(S)$ would transit to a reaction state $q$. The state $q^{dump}$ denotes the situation when an event $\sigma \in \Sigma_{o}$, which is not defined at the state $q \in Q_{b}$ in $B = \mathscr{P}_{\Sigma_{o}}(G||S)$, occurs at the state $q$ in $BPS(S)$. The initial state of $BPS(S)$ is thus the initial control state, denoted by $q_{bps}^{init} = (q_{b}^{init})^{com}$. Next, we explain the definition of $\xi_{bps}$. 
Case 1 adds any feasible control command $\gamma$ that can be issued at any control state $q^{com}$ to preserve the control equivalence. The criteria for adding $\gamma \in \Gamma$ at the state $q^{com}$ contains two conditions: 1) The sending of $\gamma$ should make sure that all the events in $En_{B}(q)$ would occur at the plant $G$ once $\gamma$ is received, denoted by the condition $\mathcal{C}_{1} := En_{B}(q) \subseteq \gamma$; 2) According to the way of constructing $B = \mathscr{P}_{\Sigma_{o}}(G||S)$, any state $q \in Q_{b}$ such that $(\exists t \in \Sigma_{o}^{*})\xi_{b}(q_{b}^{init}, t) = q$ already contains all the possible estimated states of the plant $G$ w.r.t. the observation sequence $t$. Henceforth, for any possible plant state $q_{g}$ in the state $q$, the sending of $\gamma$ should make sure that any event that might be executed at the state $q_{g}$ under $\gamma$ would not go beyond $En_{B}(q)$, denoted by the condition $\mathcal{C}_{2} := (\forall (q_{g},q_{s}) \in q)En_{G}(q_{g}) \cap \gamma \subseteq En_{B}(q)$. The conditions $\mathcal{C}_{1}$ and $\mathcal{C}_{2}$ together enforce that at the control state $q^{com}$, any control command $\gamma$ satisfying these two conditions would enable the plant to execute exactly those events in $En_{B}(q)$. Case 2 and Case 3 retain all the transitions defined in $B$, similar to the construction of $BT(S)$ in Section \ref{subsec:Component models}.
Next, we explain why we add Case 4 and Case 5. Our goal is to construct a structure to include all the control equivalent bipartite supervisors. For any supervisor $S' = (Q_{s'}, \Sigma, \xi_{s'}, q_{s'}^{init})$, at any reaction state $q \in Q_{s'}$ of its bipartite version $BT(S')$, all the events in the control command issued at the state $q^{com}$ should be defined. Since Case 1 has already added all the possible control commands that ensure control equivalence, our basic idea is: 
\begin{enumerate}[1.]
\setlength{\itemsep}{3pt}
\setlength{\parsep}{0pt}
\setlength{\parskip}{0pt}
    \item Firstly, for any reaction state $q \in Q_{b}$ in $BPS(S)$, we carry out Case 4 and Case 5 to complete all the transitions labelled by events in $\Sigma$ that are not defined at the state $q \in Q_{b}$ in $B$. The completed unobservable events would lead to self-loop transitions. The completed observable events would result in transitions to the state $q^{dump}$, where any control command is defined. Since these completed observable events would not occur at all under control commands defined at the control state $q^{com}$, the control equivalence would not be violated no matter which command is issued at the state $q^{dump}$.
    \item Then we use $CE$ to refine the above-constructed structure to get all bipartite control equivalent supervisors, which would be done in the later Section \ref{subsec:Structure refinement}. 
\end{enumerate}
In Case 6, since $BPS(S)$ has transited from some state $q \in Q_{b}$ to the state $q^{dump}$, we can safely add self-loop transitions labelled by the events in $\Gamma$ at the state $q^{dump}$ without affecting the control equivalence. We also add self-loop transitions labelled by the events in $\Sigma$ at the state $q^{dump}$ for the later structure refinement.

\begin{figure}[htbp]
\begin{center}
\includegraphics[height=3.7cm]{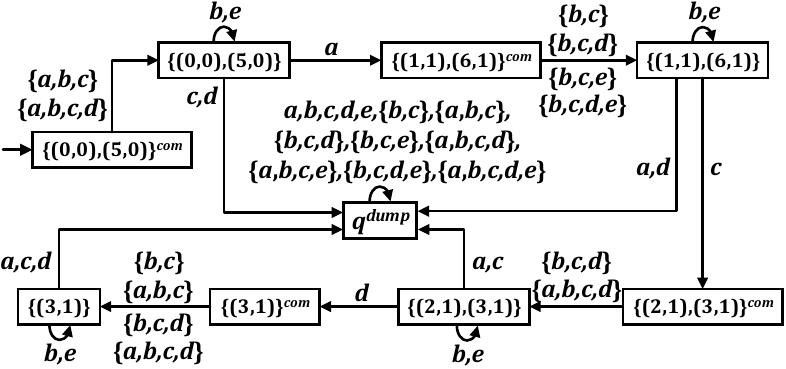}   
\caption{Bipartite behavior-preserving structure $BPS(S)$}
\label{fig:BPS}
\end{center}        
\end{figure}

\textbf{Example V.2} Based on $B$ shown in Fig. \ref{fig:B}, the constructed $BPS(S)$ is illustrated in Fig. \ref{fig:BPS}. At the initial control state $\{(0,0),(5,0)\}^{com}$, 1) according to $\mathcal{C}_{1}$ of Case 1, we have $En_{B}(\{(0,0),(5,0)\}) = \{a,b\} \subseteq \gamma$, and 2) according to $\mathcal{C}_{2}$ of Case 1, we have $En_{G}(0) = \{a,b,e\} \cap \gamma \subseteq En_{B}(\{(0,0),(5,0)\}) = \{a,b\}$ and $En_{G}(5) = \{a\} \cap \gamma \subseteq En_{B}(\{(0,0),(5,0)\}) = \{a,b\}$. Thus, the control commands satisfying $\mathcal{C}_{1}$ and $\mathcal{C}_{2}$ are $\{a,b,c\}$ and $\{a,b,c,d\}$. Hence, there are two transitions labelled by $\{a,b,c\}$ and $\{a,b,c,d\}$ from the initial state to the reaction state $\{(0,0),(5,0)\}$. At the state $\{(0,0),(5,0)\}$, according to Case 2 and Case 3, there are two transitions labelled by event $b$, which is a self-loop, and event $a$, which leads to the state $(\xi_{b}(\{(0,0),(5,0)\}, a)^{com} = \{(1,1),(6,1)\}^{com}$. According to Case 4, a transition labelled by unobservable event $e$ is added at the state $\{(0,0),(5,0)\}$. According to Case 5, two transitions labelled by observable events $c$ and $d$ are added at the state $\{(0,0),(5,0)\}$, which lead to the state $q^{dump}$. We can check that $c$ and $d$ would not occur under the initial command $\{a,b,c\}$ and $\{a,b,c,d\}$. 

\subsection{Structure refinement}
\label{subsec:Structure refinement}

By construction, although $BPS(S)$ contains all the feasible control commands that could ensure the control equivalence upon each observation, it does not exactly encode all the control equivalent bipartite supervisors due to the completion operations in Case 4-6 of $\xi_{bps}$. Notice that $CE$ encodes all the bipartite supervisors\footnote{By construction, at the state $q_{ce}^{init}$, any control command $\gamma \in \Gamma$ is defined and would lead to the state $q^{\gamma}$, where only events in $\gamma$ are defined. Any unobservable event in $\gamma$ would lead to a self-loop and any observable event in $\gamma$ would lead to the initial state $q_{ce}^{init}$.}. Since $BPS(S)$ and $CE$ have the same set of events, we know that all transitions are forced to be synchronized in $BPS(S)||CE$. Thus, we carry out the refinement on $BPS(S)$ by computing $BPS(S)||CE$.
We call $BPS(S)||CE$ the bipartite behavior-preserving command-nondeterministic\footnote{$BPNS(S)$ is deterministic, but command non-deterministic because more than one different control command may be defined at each control state.} supervisor, denoted by
$BPNS(S) = BPS(S) || CE = (Q_{bpns}, \Sigma_{bpns} = \Sigma \cup \Gamma, \xi_{bpns}, q_{bpns}^{init})$, where $Q_{bpns} = (Q_{b} \cup Q_{b}^{com} \cup \{q^{dump}\}) \times Q_{ce} = (Q_{b} \cup Q_{b}^{com} \cup \{q^{dump}\}) \times (\{q^{\gamma}|\gamma \in \Gamma\} \cup \{q_{ce}^{init}\})$. According to the structure of $BPS(S)$ and $CE$, we know that $Q_{bpns} = ((Q_{b} \cup \{q^{dump}\}) \times \{q^{\gamma}|\gamma \in \Gamma\}) \dot{\cup} ((Q_{b}^{com} \cup \{q^{dump}\}) \times \{q_{ce}^{init}\})$. Thus, we have $|Q_{bpns}| \leq (2^{|Q| \times |Q_{s}|} + 1) \times |\Gamma| + 2^{|Q| \times |Q_{s}|} + 1 = (2^{|Q| \times |Q_{s}|} + 1)(|\Gamma| + 1)$. 
For convenience, we call $Q_{bpns}^{rea} := (Q_{b} \cup \{q^{dump}\}) \times \{q^{\gamma}|\gamma \in \Gamma\}$ the set of reaction states since any event, if defined at these states, belongs to $\Sigma$, and we call $Q_{bpns}^{com} := (Q_{b}^{com} \cup \{q^{dump}\}) \times \{q_{ce}^{init}\}$ the set of control states since any event, if defined at these states, belongs to $\Gamma$. Thus, $Q_{bpns} = Q_{bpns}^{rea} \dot{\cup} Q_{bpns}^{com}$.

\begin{figure*}[htbp]
\begin{center}
\includegraphics[height=6cm]{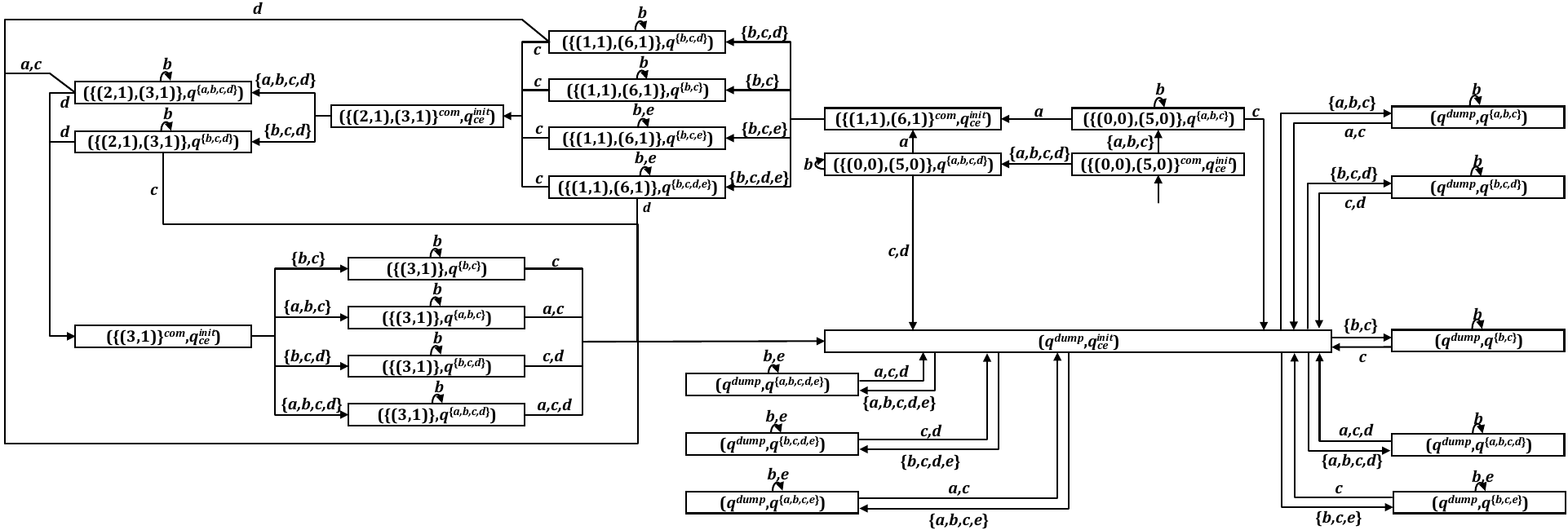}   
\caption{Bipartite behavior-preserving command-nondeterministic supervisor $BPNS(S)$}
\label{fig:BPNS}
\end{center}        
\end{figure*}

\textbf{Example V.3} Based on $BPS(S)$ shown in Fig. \ref{fig:BPS} and $CE$ shown in Fig. \ref{fig:CE_CEA}. (a), the computed $BPNS(S)$ is illustrated in Fig. \ref{fig:BPNS}. At the initial control state $(\{(0,0),(5,0)\}^{com},q_{ce}^{init})$, two control commands $\{a,b,c\}$ and $\{a,b,c,d\}$ are defined, which means that a control equivalent supervisor could issue either $\{a,b,c\}$ or $\{a,b,c,d\}$ when the system initiates. If $\{a,b,c\}$ is issued, then $BPNS(S)$ would transit to state $(\{(0,0),(5,0)\},q^{\{a,b,c\}})$, where according to the structure of a bipartite supervisor, unobservable event $b$ is a self-loop transition, and observable events $a$ and $c$ lead to two control states $(\{(1,1),(6,1)\}^{com},q_{ce}^{init})$ and $(q^{dump},q_{ce}^{init})$. 
Note that at the control state $(q^{dump},q_{ce}^{init})$, any command could be issued without violating the control equivalence since the event $c$ leading to $(q^{dump},q_{ce}^{init})$ would never occur under the issued initial command $\{a,b,c\}$ according to the structure of $G$. 

\textbf{Lemma V.1:} Given $G$ and $S$, for any supervisor $S'$, we have $L(G||S) = L(G||S')$ iff $L(\mathscr{P}_{\Sigma_{o}}(G||S)) = L(\mathscr{P}_{\Sigma_{o}}(G||S'))$.

\emph{Proof:} See Appendix \ref{appendix: Lemma V.1}. \hfill $\blacksquare$

\textbf{Proposition V.1:} Given $G$ and $S$, for any supervisor $S' = (Q_{s'}, \Sigma, \xi_{s'}, q_{s'}^{init})$ such that $L(G||S) = L(G||S')$, we have $L(BT(S')) \subseteq L(BPNS(S))$.


\emph{Proof:} See Appendix \ref{appendix: Proposition V.1}. \hfill $\blacksquare$

\textbf{Proposition V.2:} Given $G$ and $S$, for any supervisor $S' = (Q_{s'}, \Sigma, \xi_{s'}, q_{s'}^{init})$ such that $L(G||S) \neq L(G||S')$, we have $L(BT(S')) \not\subseteq L(BPNS(S))$.

\emph{Proof:} See Appendix \ref{appendix: Proposition V.2}. \hfill $\blacksquare$ 

In the following text, we shall denote by $\mathscr{S}$ the set of all the supervisors (satisfying controllability and observability), and $\mathscr{S}_{e}(S) := \{S' \in \mathscr{S} |L(G||S) = L(G||S')\}$ the set of supervisors that are control equivalent to $S$.

\textbf{Theorem V.1:} $\bigcup\limits_{S' \in \mathscr{S}_{e}(S)}L(BT(S')) = L(BPNS(S))$.

\emph{Proof:} See Appendix \ref{appendix: Theorem V.1}. \hfill $\blacksquare$ 

Based on \textbf{Theorem V.1}, $BPNS(S)$ exactly encodes all the control equivalent bipartite supervisors.


\section{Synthesis of fortified supervisors}
\label{sec:Synthesis of Fortified Supervisors Against Covert Actuator Attackers}

In this section, we introduce how to synthesize fortified supervisors based on $BPNS(S)$.

\subsection{Covert damage strings identification}
\label{subsec:Damage strings encoding}

To maintain control equivalence, any fortified supervisor cannot affect what it will observe from the execution of the closed-loop system. Thus, to find fortified supervisors from $BPNS(S)$, we could only prune those illegal control commands, which requires us to first identify the covert damage strings. However, $BPNS(S)$ only contains the control equivalent supervisors in the absence of attacks. Hence, we need to encode the actuator attack effects into $BPNS(S)$ to build the model under attack, denoted by $BPNS^{A}(S) = (Q_{bpns}^{a}, \Sigma_{bpns}^{a}, \xi_{bpns}^{a}, q_{bpns}^{a,init})$, where 
\begin{itemize}
\setlength{\itemsep}{3pt}
\setlength{\parsep}{0pt}
\setlength{\parskip}{0pt}
    \item $Q_{bpns}^{a} = Q_{bpns} \cup \{q_{bpns}^{detect}\} = Q_{bpns}^{rea} \cup Q_{bpns}^{com} \cup \{q_{bpns}^{detect}\}$
    \item $q_{bpns}^{a,init} = q_{bpns}^{init}$
    \item $\Sigma_{bpns}^{a} = \Sigma \cup \Gamma$
    \item $\xi_{bpns}^{a}$ is defined as:
    \begin{enumerate}[1.]
    \setlength{\itemsep}{3pt}
    \setlength{\parsep}{0pt}
    \setlength{\parskip}{0pt}
        \item $(\forall q, q' \in Q_{bpns}^{a})(\forall \sigma \in \Sigma \cup \Gamma) \xi_{bpns}(q, \sigma) = q' \Rightarrow \xi_{bpns}^{a}(q, \sigma) = q'$ 
        \item $(\forall q \in Q_{bpns}^{rea})(\forall \sigma \in \Sigma_{c,a} \cap \Sigma_{uo}) \neg\xi_{bpns}(q, \sigma)! \Rightarrow \xi_{bpns}^{a}(q, \sigma) = q$ 
        \item $(\forall q \in Q_{bpns}^{rea})(\forall \sigma \in \Sigma_{o}) \neg\xi_{bpns}(q, \sigma)! \Rightarrow \xi_{bpns}^{a}(q, \sigma) = q_{bpns}^{detect}$
    \end{enumerate}
\end{itemize}
The construction procedure of $BPNS^{A}(S)$ from $BPNS(S)$ is similar to that of generating $BT(S)^{A}$ from $BT(S)$ in Section \ref{subsubsec:Supervisor}. When an unexpected observation is received (Case 3), $BPNS^{A}(S)$ reaches the state $q_{bpns}^{detect}$. We have the following.

\textbf{Proposition VI.1:} Given $G$ and $S$, for any supervisor $S' = (Q_{s'}, \Sigma, \xi_{s'}, q_{s'}^{init})$ such that $L(G||S) = L(G||S')$, we have $L(BT(S')^{A}) \subseteq L(BPNS^{A}(S))$. 

\emph{Proof:} See Appendix \ref{appendix: Proposition VI.1}. \hfill $\blacksquare$

\textbf{Theorem VI.1:} $\bigcup\limits_{S' \in \mathscr{S}_{e}(S)}L(BT(S')^{A}) = L(BPNS^{A}(S))$. 

\emph{Proof:} See Appendix \ref{appendix: Theorem VI.1}. \hfill $\blacksquare$

\begin{figure*}[htbp]
\begin{center}
\includegraphics[height=6cm]{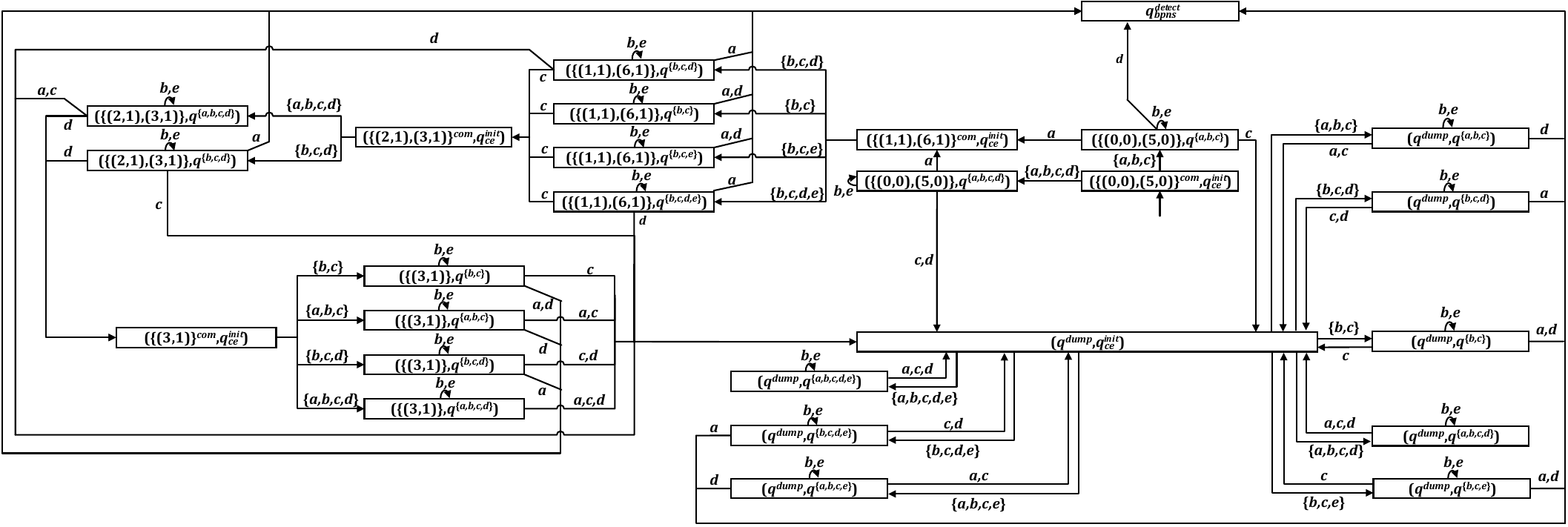}   
\caption{Bipartite behavior-preserving command-nondeterministic supervisor under attack $BPNS^{A}(S)$}
\label{fig:BPNSA}
\end{center}        
\end{figure*}

\textbf{Example VI.1} Based on $BPNS(S)$ shown in Fig. \ref{fig:BPNS}, the constructed $BPNS^{A}(S)$ is illustrated in Fig. \ref{fig:BPNSA}. 
We take the state $(\{(0,0),(5,0)\},q^{\{a,b,c\}})$ as an instance to explain how to construct $BPNS^{A}(S)$ based on $BPNS(S)$. According to Case 2 of $\xi_{bpns}^{a}$, the transition labelled by the unobservable but attackable event $e$ is added, which is a self-loop. According to Case 3 of $\xi_{bpns}^{a}$, the transition labelled by the observable event $d$, which is not defined in $BPNS(S)$, is added and leads to the state $q_{bpns}^{detect}$.


Next, we shall synthesize a structure to encode all the covert damage strings, where each one works for at least one control equivalent supervisor. The procedure is as follows:

\noindent \textbf{Procedure 1:}
\begin{enumerate}[1.]
\setlength{\itemsep}{3pt}
\setlength{\parsep}{0pt}
\setlength{\parskip}{0pt}
    \item Input: $G$, $CE^{A}$, $BPNS^{A}(S)$. In the constructed supervisory control problem at step 4, where the event set is $\Sigma \cup \Gamma$, events in $\Sigma_{c,a}$ are controllable, and events in $\Sigma_{o,a} \cup \Gamma$ are observable.
    \item Compute $\mathcal{P} = G||CE^{A}||BPNS^{A}(S) = (Q_{\mathcal{P}}, \Sigma_{\mathcal{P}}, \xi_{\mathcal{P}}, q_{\mathcal{P}}^{init}, Q_{\mathcal{P},m})$, where $Q_{\mathcal{P},m} = Q_{d} \times Q_{ce}^{a} \times Q_{bpns}^{a}$.  
    \item Generate $\mathcal{P}_{r} = \mathcal{P}^{|Q_{\mathcal{P}} - Q_{bad}}$, where $Q_{bad} = \{(q, q_{ce}^{a}, q_{bpns}^{a}) \in Q_{\mathcal{P}}|q_{bpns}^{a} = q_{bpns}^{detect}\}$.
    \item Solve a BSCOP where the plant is $\mathcal{P}$, the legal language is $L(\mathcal{P}_{r})$, and the control constraint is $\mathscr{C}_{ac}  = (\Sigma_{c,a}, \Sigma_{o,a} \cup \Gamma)$. The synthesized supremal solution is denoted as $\hat{\mathcal{A}} = (Q_{\hat{a}}, \Sigma_{\hat{a}}, \xi_{\hat{a}}, q_{\hat{a}}^{init})$.
    \item Output: $\hat{\mathcal{A}}$.
\end{enumerate}
We briefly explain \textbf{Procedure 1}. In Step 2, we generate a new plant $\mathcal{P} = G||CE^{A}||BPNS^{A}(S)$. In Step 3, we generate $\mathcal{P}_{r}$ from $\mathcal{P}$ by removing states in $Q_{bad}$, where the covertness is broken, denoted by $q_{bpns}^{a} = q_{bpns}^{detect}$. 
In Step 4, we construct a BSCOP by treating $\mathcal{P}$ as the plant and $L(\mathcal{P}_{r})$ as the legal language. Based on the assumption $\Sigma_{c,a} \subseteq \Sigma_{o,a}$, we have $\Sigma_{c,a} \subseteq \Sigma_{o,a} \cup \Gamma$. Thus, the supremal solution ${\hat{\mathcal{A}}}$ exists.
We also remark that a resilient supervisor should prevent damage infliction against any covert actuator attack. Thus, we should find covert damage strings that could be used by damage-reachable actuator attackers. Hence, in Step 4, we only need to deal with a BSCOP instead of its nonblocking version \cite{CL99}. We will show the correctness later.

In the following text, the set of covert and damage-reachable actuator attackers against the supervisor $S'$ is denoted as $\mathscr{A}(S')$.

\textbf{Proposition VI.2:} Given $G$ and $S$, for any supervisor $S'$ such that $L(G||S) = L(G||S')$ and any attacker $\mathcal{A} \in \mathscr{A}(S')$, we have $L(G||CE^{A}||BT(S')^{A}||\mathcal{A}) \subseteq L(G||CE^{A}||BPNS^{A}(S)||\hat{\mathcal{A}})$. 

\emph{Proof:} See Appendix \ref{appendix: Proposition VI.2}. \hfill $\blacksquare$

\textbf{Corollary VI.1:} Given $G$ and $S$, for any supervisor $S'$ such that $L(G||S) = L(G||S')$ and any attacker $\mathcal{A} \in \mathscr{A}(S')$, we have $L_{m}(G||CE^{A}||BT(S')^{A}||\mathcal{A}) \subseteq L_{m}(G||CE^{A}||BPNS^{A}(S)||\hat{\mathcal{A}})$.

\emph{Proof:} Based on \textbf{Proposition VI.2}, it holds that $L(G||CE^{A}||BT(S')^{A}||\mathcal{A}) ||L_{m}(G) \subseteq L(G||CE^{A}||BPNS^{A}(S)||\hat{\mathcal{A}}) || L_{m}(G)$. Since $L_{m}(G||CE^{A}||BT(S')^{A}||\mathcal{A}) = L(G||CE^{A}||BT(S')^{A}||\mathcal{A}) ||L_{m}(G)$ and $L_{m}(G||CE^{A}||BPNS^{A}(S)||\hat{\mathcal{A}}) = L(G||CE^{A}||BPNS^{A}(S)||\hat{\mathcal{A}}) || L_{m}(G)$, the proof is completed. \hfill $\blacksquare$

\textbf{Theorem VI.2:} $L_{m}(G||CE^{A}||BPNS^{A}(S)||\hat{\mathcal{A}}) = \bigcup\limits_{S' \in \mathscr{S}_{e}(S)}\bigcup\limits_{\mathcal{A} \in \mathscr{A}(S')}L_{m}(G||CE^{A}||BT(S')^{A}||\mathcal{A})$

\emph{Proof:} See Appendix \ref{appendix: Theorem VI.2}. \hfill $\blacksquare$

\textbf{Theorem VI.2} implies that $L_{m}(G||CE^{A}||BPNS^{A}(S)||\hat{\mathcal{A}})$ encodes all the covert damage strings, where each one works for at least one control equivalent supervisor. \textbf{Theorem VI.2} also implies that $L_{m}(G||CE^{A}||BPNS^{A}(S)||\hat{\mathcal{A}}) \neq \varnothing$ since the original non-resilient supervisor $S \in \mathscr{S}_{e}(S)$.

\begin{figure}[htbp]
\begin{center}
\includegraphics[height=4.5cm]{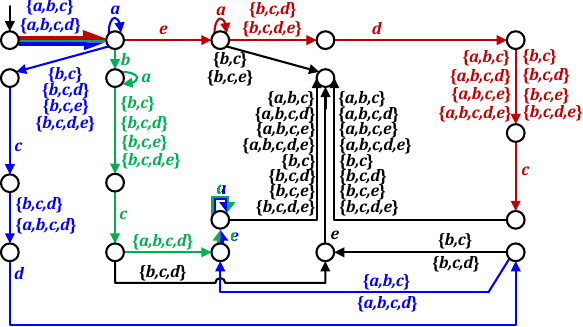}   
\caption{The synthesized $\hat{\mathcal{A}}$}
\label{fig:A}
\end{center}        
\end{figure}

\textbf{Example VI.2} Based on $G$, $CE^{A}$ and $BPNS^{A}(S)$ shown in Fig. \ref{fig:G_S_BTS_BTSA}. (a), Fig. \ref{fig:CE_CEA}. (b) and Fig. \ref{fig:BPNSA}, respectively, the synthesized $\hat{\mathcal{A}}$ by adopting \textbf{Procedure 1} is illustrated in Fig. \ref{fig:A}. 
It can be checked that $\hat{\mathcal{A}}$ encodes three kinds of damage strings (marked in red, green and blue) that can be taken use of by the attacker. We take the red part as an instance. After observing the initial control command $\{a,b,c\}$ or $\{a,b,c,d\}$, the attacker can carry out the enablement attack to enable the execution of unobservable event $e$, which results in the reuse of initial control command and event $a$ is executed. After that, if the supervisor issues a control command containing event $d$, that is, command $\{b,c,d\}$ or $\{b,c,d,e\}$, then event $d$ would be executed, triggering the command sending by the supervisor. Finally, event $c$ is executed, causing the damage infliction.

\subsection{Illegal control commands pruning}
\label{subsec:Illegal control commands pruning}

As stated before, to obtain the fortified supervisors from $BPNS(S)$, we need to prune inappropriate transitions labelled by control commands in $\Gamma$, which are controllable to the supervisor. Now, since $L_{m}(G||CE^{A}||BPNS^{A}(S)||\hat{\mathcal{A}})$ encodes all the covert damage strings based on \textbf{Theorem VI.2}, the intuitive idea of our methodology is to extract the fortified supervisors under attack by treating $BPNS^{A}(S)$ as a plant and then performing the synthesis with the guidance of $L_{m}(G||CE^{A}||BPNS^{A}(S)||\hat{\mathcal{A}})$.
The detailed methodology is presented as follows. 

\noindent \textbf{Procedure 2:}

\begin{enumerate}[1.]
\setlength{\itemsep}{3pt}
\setlength{\parsep}{0pt}
\setlength{\parskip}{0pt}
    \item Input: $G$, $CE^{A}$, $BPNS^{A}(S)$, $\hat{\mathcal{A}}$. In the constructed supervisory control problem at step 4, where the event set is $\Sigma \cup \Gamma$, events in $\Gamma$ are controllable, and events in $\Sigma_{o} \cup \Gamma$ are observable.
    \item Compute $\mathcal{P} = G||CE^{A}||BPNS^{A}(S)||\hat{\mathcal{A}} = (Q_{\mathcal{P}}, \Sigma_{\mathcal{P}} = \Sigma \cup \Gamma, \xi_{\mathcal{P}}, q_{\mathcal{P}}^{init}, Q_{\mathcal{P},m})$.
    \item Construct $\mathcal{P}_{r} = (Q_{\mathcal{P}_{r}}, \Sigma_{\mathcal{P}_{r}}, \xi_{\mathcal{P}_{r}}, q_{\mathcal{P}_{r}}^{init})$ based on $\mathcal{P}$, where 
    \begin{enumerate}[a.]
    \setlength{\itemsep}{3pt}
    \setlength{\parsep}{0pt}
    \setlength{\parskip}{0pt}
        \item $Q_{\mathcal{P}_{r}} = (Q_{\mathcal{P}} - Q_{\mathcal{P},m}) \cup \{q^{dump}\}$
        \item $\Sigma_{\mathcal{P}_{r}} = \Sigma \cup \Gamma$
        \item $q_{\mathcal{P}_{r}}^{init} = q_{\mathcal{P}}^{init}$
        \item $\xi_{\mathcal{P}_{r}}$ is defined as:
        \begin{enumerate}[i.]
        \setlength{\itemsep}{3pt}
        \setlength{\parsep}{0pt}
        \setlength{\parskip}{0pt}
            \item $(\forall q, q' \in Q_{\mathcal{P}} - Q_{\mathcal{P},m})(\forall \sigma \in \Sigma \cup \Gamma)\xi_{\mathcal{P}}(q, \sigma) = q' \Rightarrow \xi_{\mathcal{P}_{r}}(q, \sigma) = q'$
            \item $(\forall q \in Q_{\mathcal{P}} - Q_{\mathcal{P},m})(\forall \sigma \in \Sigma \cup \Gamma)\neg\xi_{\mathcal{P}}(q, \sigma)! \Rightarrow \xi_{\mathcal{P}_{r}}(q, \sigma) = q^{dump}$ 
            \item $(\forall \sigma \in \Sigma \cup \Gamma)\xi_{\mathcal{P}_{r}}(q^{dump}, \sigma) = q^{dump}$
        \end{enumerate}
    \end{enumerate}
    \item Solve a BSCOP where the plant is $BPNS^{A}(S)$, the legal language is $L(\mathcal{P}_{r})$, and the control constraint is $(\Gamma, \Sigma_{o} \cup \Gamma)$. The synthesized supremal solution is denoted as $S_{0}^{A} = (Q_{S_{0}^{A}}, \Sigma_{S_{0}^{A}} = \Sigma \cup \Gamma, \xi_{S_{0}^{A}}, q_{S_{0}^{A}}^{init})$.
    \item Output: $S_{0}^{A}$.
\end{enumerate}  
In Step 2, we compute $\mathcal{P} = G||CE^{A}||BPNS^{A}(S)||\hat{\mathcal{A}}$, whose marked behavior encodes all the covert damage strings. Notice that $L_{m}(\mathcal{P}) \neq \varnothing$ and $Q_{\mathcal{P},m} \neq \varnothing$ according to \textbf{Theorem VI.2}.
In Step 3, we construct $\mathcal{P}_{r}$ based on $\mathcal{P}$. The state set of $\mathcal{P}_{r}$ is constructed by removing the set of marker states of $\mathcal{P}$ and adding a new state $q^{dump}$, as shown in Step 3.a. $\xi_{\mathcal{P}_{r}}$ is defined as follows: 1) for any two states that have not been removed, the transitions between them defined in $\mathcal{P}$ are retained in $\mathcal{P}_{r}$, as shown in Step 3.d.i, 2) for any state $q$ that has not been removed, we complete the transitions that are not defined at state $q$ in $\mathcal{P}$, which would lead to the newly added state $q^{dump}$, as shown in Step 3.d.ii, and 3) all the transitions in $\Sigma \cup \Gamma$ are defined at the state $q^{dump}$ in Step 3.d.iii. Since we are only supposed to forbid the execution of the strings that might result in damage infliction, we carry out Step 3.d.ii and Step 3.d.iii such that $\mathcal{P}_{r}$ specifies all the legal strings. 
It is noteworthy that, although Step 3.d.ii completes the transitions that are not defined in $\mathcal{P}$ for those states in $Q_{\mathcal{P}} - Q_{\mathcal{P},m}$ and Step 3.d.iii adds self-loops labelled as events in $\Sigma \cup \Gamma$ for the state $q^{dump}$, $\mathcal{P}_{r}$ is not a complete automaton because when any state $q \in Q_{\mathcal{P},m}$ is removed from $\mathcal{P}_{r}$, all the transitions attached to this state $q$ are also removed.
In Step 4, we construct a BSCOP by treating $BPNS^{A}(S)$ as the plant and $L(\mathcal{P}_{r})$ as the legal language. Since $\Gamma \subseteq \Sigma_{o} \cup \Gamma$, we could always synthesize the supremal solution $S_{0}^{A}$. We remark that the attacked versions of those fortified supervisors in $BPNS(S)$ has been included in $S_{0}^{A}$. We will show the correctness later.

\textbf{Remark VI.1:} Since $S_{0}^{A}$ is synthesized by treating $BPNS^{A}(S)$ as the plant in Step 4 of \textbf{Procedure 2}, we consider the case where $S_{0}^{A}$ satisfies that $L(S_{0}^{A}) \subseteq L(BPNS^{A}(S))$,  following the standard notion of controllability and observability \cite{WMW10} over the control constraint $(\Gamma, \Sigma_{o} \cup \Gamma)$ w.r.t. the plant $BPNS^{A}(S)$. Without loss of generality, any event in $\Sigma_{uo}$, if defined, is a self-loop transition in $S_{0}^{A}$. Thus, $S_{0}^{A}$ is a bipartite structure\footnote{If we follow our definition of a supervisor and synthesize $S_{0}^{A}$, we could always update $S_{0}^{A} := BPNS^{A}(S)||S_{0}^{A}$ to generate a bipartite structure $S_{0}^{A}$ with $L(S_{0}^{A}) \subseteq L(BPNS^{A}(S))$.} similar to $BPNS^{A}(S)$.

\begin{figure}[htbp]
\begin{center}
\includegraphics[height=4.7cm]{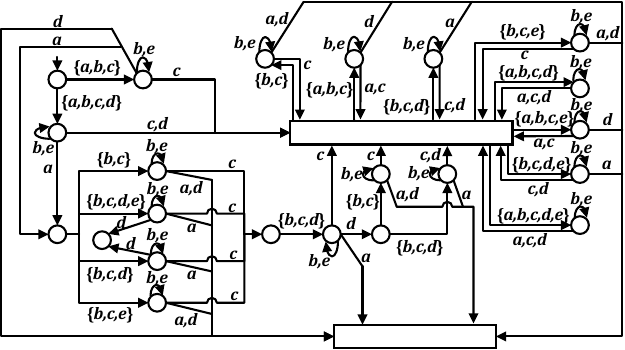}   
\caption{The synthesized $S_{0}^{A}$}
\label{fig:S0A}
\end{center}        
\end{figure}

\textbf{Example VI.3} Based on $G$, $CE^{A}$, $BPNS^{A}(S)$ and $\hat{\mathcal{A}}$ shown in Fig. \ref{fig:G_S_BTS_BTSA}. (a), Fig. \ref{fig:CE_CEA}. (b), Fig. \ref{fig:BPNSA} and Fig. \ref{fig:A}, respectively, the synthesized $S_{0}^{A}$ by adopting \textbf{Procedure 2} is illustrated in Fig. \ref{fig:S0A}. 
Compared with $BPNS^{A}(S)$ shown in Fig. \ref{fig:BPNSA}, there are several control commands removed at some control states in $S_{0}^{A}$.
For example, after the occurrence of the sequence $\{a,b,c\}/\{a,b,c,d\} \rightarrow a \rightarrow \{b,c,d\}/\{b,c,d,e\} \rightarrow d$, the plant now might execute the string $ead$ due to the enablement attack of unobservable event $e$. In this case, any control command cannot be issued because the string $eadc$ would cause damage infliction and uncontrollable event $c$ is contained in any control command. This case corresponds to the defense strategy against the damage string marked in red, which is shown in \textbf{Example VI.2}.

\subsection{Fortified supervisors synthesis}
\label{subsec:Generation of Fortified Supervisors}

The synthesized $S_{0}^{A}$ contains the attacked version of fortified (resilient and control equivalent) supervisors. Since the goal is to synthesize the non-attacked version of fortified supervisors, we transform $S_{0}^{A}$ to the version in the absence of attacks, denoted as $S_{0} = (Q_{S_{0}}, \Sigma_{S_{0}}, \xi_{S_{0}}, q_{S_{0}}^{init})$, where 
\begin{itemize}
\setlength{\itemsep}{3pt}
\setlength{\parsep}{0pt}
\setlength{\parskip}{0pt}
    \item $Q_{S_{0}} = Q_{S_{0}^{A}}$
    \item $q_{S_{0}}^{init} = q_{S_{0}^{A}}^{init}$
    \item $\Sigma_{S_{0}} = \Sigma \cup \Gamma$
    \item $\xi_{S_{0}}$ is defined as:
    \begin{enumerate}[1.]
    \setlength{\itemsep}{3pt}
    \setlength{\parsep}{0pt}
    \setlength{\parskip}{0pt}
        \item $(\forall q, q' \in Q_{S_{0}})(\forall \gamma \in \Gamma)\xi_{S_{0}^{A}}(q, \gamma) = q' \Rightarrow \xi_{S_{0}}(q, \gamma) = q'$
        \item $(\forall q, q' \in Q_{S_{0}})(\forall \gamma \in \Gamma)(\forall \sigma \in \gamma \cap \Sigma_{uo})\xi_{S_{0}^{A}}(q, \gamma) = q' \Rightarrow \xi_{S_{0}}(q', \sigma) = q'$
        \item $(\forall q, q', q'' \in Q_{S_{0}})(\forall \gamma \in \Gamma)(\forall \sigma \in \gamma \cap \Sigma_{o})\xi_{S_{0}^{A}}(q, \gamma) =\\ q' \wedge \xi_{S_{0}^{A}}(q', \sigma) = q'' \Rightarrow \xi_{S_{0}}(q', \sigma) = q''$
    \end{enumerate}
\end{itemize}
Briefly speaking, 1) we retain all the transitions labelled by events in $\Gamma$ that are originally defined in $S_{0}^{A}$, as shown in Case 1, 2) for any state $q'$ such that there exists a transition $\xi_{S_{0}^{A}}(q, \gamma) = q'$, we retain the transition labelled by any event in $\gamma \cap \Sigma_{uo}$ ($\gamma \cap \Sigma_{o}$, respectively), which is a self-loop (leads to a new state $q''$, respectively), as shown in Case 2 (Case 3, respectively). 
Then we generate the automaton $Ac(S_{0})$. For convenience, in the rest, we shall refer to $Ac(S_{0})$ whenever we talk about $S_{0}$.
By Remark VI.1, $S_{0}$ is a bipartite structure and the state set of $S_{0}$ could be divided into two disjoint sets $Q_{S_{0}} = Q_{S_{0}}^{rea} \dot{\cup} Q_{S_{0}}^{com}$, where $Q_{S_{0}}^{rea}$ is the set of reaction states and $Q_{S_{0}}^{com}$ is the set of control states, satisfying that 1) at any state of $Q_{S_{0}}^{rea}$, any event in $\Gamma$ is not defined, 2) at any state of $Q_{S_{0}}^{rea}$, any event in $\Sigma_{uo}$, if defined, leads to a self-loop, and any event in $\Sigma_{o}$, if defined, would lead to a transition to a control state, 3) at any state of $Q_{S_{0}}^{com}$, any event in $\Sigma$ is not defined, and 4) at any state of $Q_{S_{0}}^{com}$, any event in $\Gamma$, if defined, would lead to a transition to a reaction state.

\begin{figure}[htbp]
\begin{center}
\includegraphics[height=4cm]{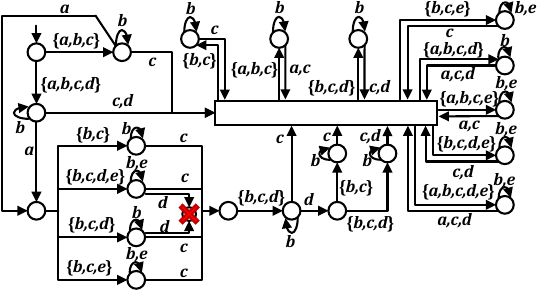}   
\caption{The transformed $S_{0}$}
\label{fig:S0}
\end{center}        
\end{figure}

\textbf{Example VI.4} Based on $S_{0}^{A}$ shown in Fig. \ref{fig:S0A}, the transformed $S_{0}$ is illustrated in Fig. \ref{fig:S0}.
We take the initial state as an instance to explain how to carry out the transformation. After the control command $\{a,b,c\}$ occurs at the initial state, according to Cases 2 and 3 of $\xi_{S_{0}}$, the transition labelled by events $a,b,c$ are retained, while the transition labelled by the event $e$ is removed.

Although inappropriate control commands leading to damage infliction have been removed in $S_{0}$, we cannot ensure $S_{0}$ exactly encodes all the fortified supervisors because it is possible that at some control state of $S_{0}$, where an observation has just been received, there is no control command defined as a result of the synthesis. Such a phenomenon violates the structure of a bipartite supervisor, where a control command must be defined at any control state according to the construction in Section \ref{subsubsec:Supervisor}. Thus, by treating $S_{0}$ as a plant, we carry out the following procedure to iteratively remove these newly created bad states until the generated structure satisfies the condition that at least a control command follows any observation.

\noindent \textbf{Procedure 3:}

\begin{enumerate}[1.]
\setlength{\itemsep}{3pt}
\setlength{\parsep}{0pt}
\setlength{\parskip}{0pt}
    \item Input: $S_{0}$. In the constructed supervisory control problem at step 6, where the event set is $\Sigma \cup \Gamma$, events in $\Gamma$ are controllable, and events in $\Sigma_{o} \cup \Gamma$ are observable.
    \item Let $k := 0$.
    \item Compute $Q_{k,del} := \{q \in Q_{S_{k}}^{com}|En_{S_{k}}(q) = \varnothing\}$. 
    \item If $Q_{k,del} = \varnothing$, then denote $FNS(S):= S_{k}$ and proceed to Step 8; otherwise, i.e., $Q_{k,del} \neq \varnothing$, then proceed to Step 5. 
    \item Construct $S_{k,r} = S_{k}^{|Q_{S_{k}} - Q_{k,del}}$.
    \item Solve a BSCOP where the plant is $S_{k}$, the legal language is $L(S_{k,r})$, and the control constraint is $(\Gamma, \Sigma_{o} \cup \Gamma)$. The synthesized supremal solution is denoted as $S_{k+1} = (Q_{S_{k+1}}, \Sigma_{S_{k+1}} = \Sigma \cup \Gamma, \xi_{S_{k+1}}, q_{S_{k+1}}^{init})$. We also denote
    $Q_{S_{k+1}} = Q_{S_{k+1}}^{rea} \dot{\cup} Q_{S_{k+1}}^{com}$, where $Q_{S_{k+1}}^{rea}$ is the set of reaction states and $Q_{S_{k+1}}^{com}$ is the set of control states\footnote{The division rule is the same as that of $Q_{S_{0}} = Q_{S_{0}}^{rea} \dot{\cup} Q_{S_{0}}^{com}$.}.
    \item Let $k \leftarrow k+1$ and proceed to Step 3.
    \item Output: $FNS(S)$.
\end{enumerate}
In Step 2, we set the counter $k$ to 0. In Step 3, taking the $k$-th iteration as an instance, we compute the set of control states in $S_{k}$, denoted by $Q_{k,del}$, where any $q \in Q_{k,del}$ satisfies that there is no control command defined, denoted by $En_{S_{k}}(q) = \varnothing$. 
In Step 4, if $Q_{k,del} = \varnothing$, then $FNS(S) := S_{k}$ is the desired structure and we output $FNS(S)$ in Step 8; otherwise, we remove $Q_{k,del}$ in $S_{k}$ to construct $S_{k,r}$ in Step 5. In Step 6, we construct a BSCOP by treating $S_{k}$ as the plant and $L(S_{k,r})$ as the legal language. Since $\Gamma \subseteq \Sigma_{o} \cup \Gamma$, we could always synthesize the supremal solution $S_{k+1}$. We name the output $FNS(S)$ of \textbf{Procedure 3} as fortified command-nondeterministic supervisor, and denote $FNS(S) = (Q_{fns}, \Sigma_{fns} = \Sigma \cup \Gamma, \xi_{fns}, q_{fns}^{init})$. In addition, we denote $Q_{fns} = Q_{fns}^{rea} \dot{\cup} Q_{fns}^{com}$, where $Q_{fns}^{rea}$ is the set of reaction states and $Q_{fns}^{com}$ is the set of control states\footnote{The division rule is the same as that of $Q_{S_{0}} = Q_{S_{0}}^{rea} \dot{\cup} Q_{S_{0}}^{com}$.}.

\textbf{Remark VI.2:} Similar to Remark VI.1, we consider the case where $S_{k+1}$ satisfies that  $L(S_{k+1}) \subseteq L(S_{k})$, following the standard notion of controllability and observability \cite{WMW10} over the control constraint $(\Gamma, \Sigma_{o} \cup \Gamma)$. Without loss of generality, any event in $\Sigma_{uo}$, if defined, is a self-loop transition in $S_{k+1}$. Thus, $S_{k+1}$ is a bipartite structure.

\textbf{Proposition VI.3:} $L(FNS(S)) \subseteq L(BPNS(S))$.

\emph{Proof:} See Appendix \ref{appendix: Proposition VI.3}. \hfill $\blacksquare$

\textbf{Theorem VI.3:} $\bigcup\limits_{S' \in \mathscr{S}_{f}(S)}L(BT(S')) = L(FNS(S))$, where $\mathscr{S}_{f}(S)$ denotes the set of fortified supervisors for $S$.

\emph{Proof:} See Appendix \ref{appendix: Theorem VI.3}. \hfill $\blacksquare$

\textbf{Example VI.5} We shall continue with $S_{0}$ shown in Fig. \ref{fig:S0}, 
\vspace{-0.2cm}
\begin{figure}[htbp]
\begin{center}
\includegraphics[height=3.4cm]{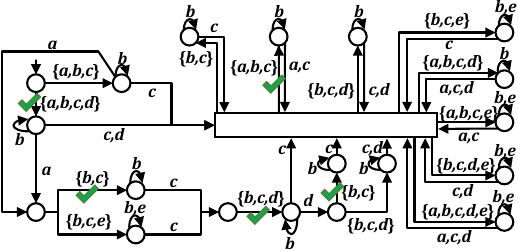}   
\caption{The synthesized $S_{1}$ ($FNS(S)$)}
\label{fig:S1}
\end{center}        
\end{figure}
\vspace{-0.2cm}
where there exists a control state (marked by a red cross) and there is no control command defined at this state. According to Step 5 of \textbf{Procedure 3}, we remove this state to generate $S_{0,r}$. By treating $S_{0}$ as the plant and $L(S_{0,r})$ as the legal language, we synthesize $S_{1}$, which is illustrated in Fig. \ref{fig:S1}. It can be checked that at least a control command is defined at any control state of $S_{1}$, which means that $Q_{1,del} = \varnothing$. Thus, the procedure terminates after the first iteration and outputs $FNS(S) := S_{1}$. 

Based on \textbf{Theorem VI.3}, $FNS(S)$ exactly encodes all the fortified supervisors. Next, we show how to extract one fortified supervisor from $FNS(S)$. We construct the following structure, denoted by $FS(S) = (Q_{fs}, \Sigma_{fs}, \xi_{fs}, q_{fs}^{init})$, where 
\begin{itemize}
\setlength{\itemsep}{3pt}
\setlength{\parsep}{0pt}
\setlength{\parskip}{0pt}
    \item $Q_{fs} = Q_{fns}$
    \item $q_{fs}^{init} = q_{fns}^{init}$
    \item $\Sigma_{fs} = \Sigma \cup \Gamma$
    \item $\xi_{fs}$ is defined as:
    \begin{enumerate}[1.]
    \setlength{\itemsep}{3pt}
    \setlength{\parsep}{0pt}
    \setlength{\parskip}{0pt}
        \item $(\forall q, q' \in Q_{fs})(\forall \sigma \in \Sigma)\xi_{fns}(q, \sigma) = q' \Rightarrow \xi_{fs}(q, \sigma) = q'$
        \item For any control state $q \in Q_{fns}^{com}$, we randomly pick a control command $\gamma \in En_{FNS(S)}(q)$ and define that: for any reaction state $q' \in Q_{fns}^{rea}$, if $\xi_{fns}(q, \gamma) = q'$, then $\xi_{fs}(q, \gamma) = q'$ and for any control command $\gamma' \in En_{FNS(S)}(q) - \{\gamma\}$, we have $\neg\xi_{fs}(q, \gamma')!$.
    \end{enumerate}
\end{itemize}
We retain all the transitions defined at any reaction state of $FNS(S)$, denoted by Case 1, and retain only one transition labelled by a control command at any control state of $FNS(S)$, denoted by Case 2. Then we generate the automaton $Ac(FS(S))$. 
For convenience, in the rest, we shall refer to $Ac(FS(S))$ whenever we talk about $FS(S)$.

\textbf{Proposition VI.4:} Given $G$ and $S$, we have $FS(S) \in \mathscr{S}_{f}(S)$.

\emph{Proof:} See Appendix \ref{appendix: Proposition VI.4}. \hfill $\blacksquare$

\textbf{Theorem VI.4:} \textbf{Problem 1} is decidable.

\emph{Proof:} To prove this result, based on \textbf{Theorem VI.3} and \textbf{Proposition VI.4}, we just need to additionally check whether \textbf{Procedure 1}, \textbf{Procedure 2}, \textbf{Procedure 3} and the extraction step could terminate within finite steps. Clearly, \textbf{Procedure 1}, \textbf{Procedure 2}, and the extraction step terminate within finite steps. For \textbf{Procedure 3}, in each iteration, since $S_{k,r}$ is generated by removing at least one control state $q$ from the plant $S_{k}$ and any unobservable event in $\Sigma_{uo}$, if defined, is a self-loop in $S_{k}$, we know that $S_{k+1}$ is a substructure of $S_{k}$. In addition, to satisfy the controllability w.r.t. the control constraint $(\Gamma, \Sigma_{o} \cup \Gamma)$, at least two states of $S_{k}$ are removed after synthesis, including the removed control state $q$ and the reaction state $q'$ where there exists $\sigma \in \Sigma_{o}$ such that $\xi_{S_{k}}(q', \sigma) = q$. Thus, \textbf{Procedure 3} would iterate Steps 3-7 for at most $\lfloor \frac{|Q_{S_{0}}|}{2} \rfloor$ times, which completes the proof. \hfill $\blacksquare$


The computational complexity of the proposed procedure depends on the complexity of three synthesis steps (\textbf{Procedure 1}, \textbf{Procedure 2} and \textbf{Procedure 3}), and the construction of $S_{0}$ from $S_{0}^{A}$. By using the synthesis approach in \cite{WMW10,WLLW18}, the complexity of \textbf{Procedure 1} is no more than $O((|\Sigma| + |\Gamma|)2^{|Q| \times |Q_{ce}^{a}| \times |Q_{bpns}^{a}|})$, the complexity of \textbf{Procedure 2} is no more than $O((|\Sigma| + |\Gamma|)2^{|Q| \times |Q_{ce}^{a}| \times |Q_{bpns}^{a}|^{2} \times |Q_{\hat{a}}|})$,
and the complexity of \textbf{Procedure 3} is no more than $O((|\Sigma| + |\Gamma|)|Q_{S_{0}}|^{2})$. The complexity of constructing $S_{0}$ from $S_{0}^{A}$ is $O((|\Sigma| + |\Gamma|)|Q_{S_{0}^{A}}|)$. Thus, the overall complexity is no more than $O((|\Sigma| + |\Gamma|)(2^{|Q| \times |Q_{ce}^{a}| \times |Q_{bpns}^{a}|^{2} \times |Q_{\hat{a}}|} + |Q_{S_{0}}|^{2}))$, where $|Q_{ce}^{a}| = |\Gamma| + 1$, $|Q_{bpns}^{a}| \leq  (2^{|Q| \times |Q_{s}|} + 1)(|\Gamma| + 1) + 1$, $|Q_{S_{0}}| = |Q_{S_{0}^{A}}| - 1$, $|Q_{S_{0}^{A}}| \leq 2^{|Q| \times |Q_{ce}^{a}| \times |Q_{bpns}^{a}|^{2} \times |Q_{\hat{a}}|}$, $|Q_{\hat{a}}| \leq 2^{|Q| \times |Q_{ce}^{a}| \times |Q_{bpns}^{a}|}$, and $|\Gamma| = 2^{|\Sigma_{c}|}$. 
In summary, the upper bound of the synthesis complexity is triply exponential w.r.t. $|Q|$, $|Q_{s}|$, and $|\Sigma_{c}|$.

\textbf{Remark VI.3}: Designing an algorithm with lower complexity for the fortified supervisor synthesis is important for practical systems. However, this is beyond the scope of this work, which focuses on presenting a technical result to show the problem of determining the existence of fortified supervisors against covert actuator attacks is decidable. In this work, we design a sound and complete procedure to assist us in proving the decidability result. Although our proposed decision process may serve as a concrete synthesis procedure to synthesize a fortified supervisor, it is just one alternative to prove this decidability result. It is possible that there exist other approaches to achieve the same goal, but we are not sure whether they are associated with lower complexities; after all, the only existing result shown in \cite{Su20}, which shows the problem of determining the existence of resilient supervisors against covert sensor attacks is decidable, also proposes a triply-exponential decision process. Thus, it is foreseen that our proposed process maybe also attached with a triply-exponential complexity. In fact, prior to embarking on any endeavor to tackle a complexity challenge, it is essential to address a fundamental question of computability: specifically, how to decide the existence of a solution. Therefore, the primary focus of this study is to demonstrate the decidability result. This significance lies in the fact that it establishes the existence of a systematic approach for obtaining correct solutions to the problem under investigation. With this decidability result, theoretical guidance is provided when we develop more efficient algorithms in the design process.

\begin{figure}[htbp]
\begin{center}
\includegraphics[height=1.4cm]{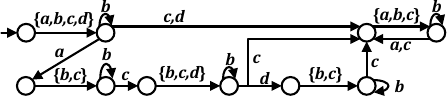}   
\caption{A fortified supervisor $FS(S)$ extracted from $FNS(S)$}
\label{fig:OS}
\end{center}        
\end{figure}

\textbf{Example VI.6} Based on $FNS(S)$ shown in Fig. \ref{fig:S1}, by choosing the control command marked by a green check mark at each control state, a fortified supervisor $FS(S)$ is extracted, which is illustrated in Fig. \ref{fig:OS}. 

\emph{Discussion:} Although the main body of this paper focuses on the decidability of the existence of fortified supervisors against covert actuator attacks, it is not difficult to extend the result to the supervisor fortification against the worst-case sensor-actuator attack. Here, we provide an alternative decision procedure. The basic idea is to transform this problem into a supervisor synthesis problem as the worst-case attack model is fixed. We can first compose the behavior-preserving structure with the plant, the command execution automaton under actuator attack, and sensor attack constraint \cite{LS20J} to construct a transformed plant and remove the damage states in this transformed plant to construct the legal language. Then we perform the synthesis together with \textbf{Procedure 3} to obtain a bipartite structure that exactly encodes all the fortified supervisors against the worst-case attack, from which we could extract one desired supervisor, completing the decision procedure.



\section{Conclusions}
\label{sec:conclusions}

In this work, we propose a sound and complete procedure to show the problem of determining the existence of fortified supervisors against covert actuator attackers is decidable. 
The limitation of this work is that we assume the system designer has the additional information that any potential attacker intends to remain covert, different from the worst-case attacker against which the resilient supervisor synthesis is not as challenging. In future works, we shall consider defending against covert sensor-actuator attackers and relax the assumption $\Sigma_{c,a} \subseteq \Sigma_{o,a}$ to study the decidability problem. Besides, developing a synthesis algorithm with lower complexity for supervisor fortification is also a pressing issue that needs to be resolved urgently.
 
\begin{appendices}
\section{Proof of Lemma V.1} 
\label{appendix: Lemma V.1} 
(If) Firstly, it can be checked that $L(\mathscr{P}_{\Sigma_{o}}(G||S)) \subseteq L(\mathscr{P}_{\Sigma_{o}}(G)||\mathscr{P}_{\Sigma_{o}}(S)) = L(\mathscr{P}_{\Sigma_{o}}(G)||S)$. Next, we prove that $L(G||S) \subseteq L(G||S')$. Thus, we need to show that for any $t \in L(G||S)$, we have $t \in L(G||S')$. Since $t \in L(G||S)$, we have $t \in L(G)$ and $t \in L(S)$. Thus, to prove $t \in L(G||S') = L(G) \cap L(S')$, we only need to show $t \in L(S')$. Since $t \in L(G||S) \subseteq L(\mathscr{P}_{\Sigma_{o}}(G||S))$, we have $t \in L(\mathscr{P}_{\Sigma_{o}}(G||S)) = L(\mathscr{P}_{\Sigma_{o}}(G||S')) \subseteq L(\mathscr{P}_{\Sigma_{o}}(G)||S') = L(\mathscr{P}_{\Sigma_{o}}(G)) \cap L(S')$, which implies that $t \in L(S')$. Thus, $L(G||S) \subseteq L(G||S')$. By the same way, we could prove that $L(G||S') \subseteq L(G||S)$. Hence, $L(G||S) = L(G||S')$.

(Only if) The necessity is straightforward. \hfill $\blacksquare$

\section{Proof of Proposition V.1} 
\label{appendix: Proposition V.1} 
Since $L(BT(S')) \subseteq L(CE)$ and $L(BPNS(S)) = L(BPS(S)) \cap L(CE)$, to prove $L(BT(S')) \subseteq L(BPNS(S))$, we only need to show that $L(BT(S')) \subseteq L(BPS(S))$. Thus, we need to prove for any $t \in L(BT(S'))$, we have $t \in L(BPS(S))$. We adopt the mathematical induction to prove this result. The base case is: $t = \varepsilon$. Clearly, $\varepsilon \in L(BT(S'))$ and $\varepsilon \in L(BPS(S))$. Next, the induction hypothesis is that: for any $t \in L(BT(S'))$, we have $t \in L(BPS(S))$, when $|t| = k$. Then we show that for any $t\sigma \in L(BT(S'))$, we have $t\sigma \in L(BPS(S))$. For convenience, we denote $BT(S') = (Q_{bs'}, \Sigma \cup \Gamma, \xi_{bs'}, q_{bs'}^{init})$. By construction, we have $L(BT(S')) \subseteq \overline{(\Gamma\Sigma_{uo}^{*}\Sigma_{o})^{*}}$, and then the verification can be divided into the following two cases:

\noindent 1. $\sigma \in \Gamma$. For convenience, we denote $\sigma = \gamma \in \Gamma$. Based on the structure of $BT(S')$, we have $t \in (\Gamma\Sigma_{uo}^{*}\Sigma_{o})^{*}$. Then we have the following two subcases:
\begin{enumerate}[1)]
\setlength{\itemsep}{3pt}
\setlength{\parsep}{0pt}
\setlength{\parskip}{0pt}
    \item $\xi_{bps}(q_{bps}^{init}, t) = q^{dump}$. Based on Case 6 of $\xi_{bps}$ of $BPS(S)$, we have $En_{BPS(S)}(\xi_{bps}(q_{bps}^{init}, t)) =  \Gamma \cup \Sigma$. Thus, it holds that $t\gamma \in L(BPS(S))$.
    \item $\xi_{bps}(q_{bps}^{init}, t) \neq q^{dump}$. We show that at the state $\xi_{bps}(q_{bps}^{init}, t)$, $\gamma$ satisfies the conditions $\mathcal{C}_{1}$ and $\mathcal{C}_{2}$ presented in Case 1 of $\xi_{bps}$. Since $L(G||S) = L(G||S')$, we have $L(B) = L(\mathscr{P}_{\Sigma_{o}}(G||S)) = L(\mathscr{P}_{\Sigma_{o}}(G||S')) \subseteq L(S')$. Thus, we have $En_{B}(\xi_{b}(q_{b}^{init}, P(t))) \subseteq En_{S'}(\xi_{s'}(q_{s'}^{init}, P(t))) = \gamma$, which satisfies $\mathcal{C}_{1}$. For $\mathcal{C}_{2}$, it requires that $(\forall (q_{g},q_{s}) \in \xi_{b}(q_{b}^{init}, P(t)))En_{G}(q_{g}) \cap En_{S'}(\xi_{s'}(q_{s'}^{init}, P(t))) \subseteq En_{B}(\xi_{b}(q_{b}^{init}, P(t)))$, which clearly holds; otherwise, we know that there exists $(q_{g},q_{s}) \in \xi_{b}(q_{b}^{init}, P(t))$ such that $En_{G}(q_{g}) \cap En_{S'}(\xi_{s'}(q_{s'}^{init}, P(t))) \not\subseteq En_{B}(\xi_{b}(q_{b}^{init}, P(t)))$, and then we have $L(\mathscr{P}_{\Sigma_o}(G||S)) \neq L(\mathscr{P}_{\Sigma_o}(G||S'))$, implying that $L(G||S) \neq L(G||S')$ based on \textbf{Lemma V.1}, which causes the contradiction.
\end{enumerate}
2. $\sigma \in \Sigma$. By construction, there exists $t_{1} \in (\Gamma\Sigma_{uo}^{*}\Sigma_{o})^{*}$, $\gamma \in \Gamma$ and $t_{2} \in (\gamma \cap \Sigma_{uo})^{*}$ such that $t = t_{1}\gamma t_{2} \in L(BPS(S))$ and $\sigma \in \gamma$. Then we have the  following two subcases:
\begin{enumerate}[1)]
\setlength{\itemsep}{3pt}
\setlength{\parsep}{0pt}
\setlength{\parskip}{0pt}
    \item $\xi_{bps}(q_{bps}^{init}, t) = q^{dump}$. Based on Case 6 of $\xi_{bps}$ of $BPS(S)$, we have $En_{BPS(S)}(\xi_{bps}(q_{bps}^{init}, t)) =  \Gamma \cup \Sigma$. Thus, it holds that $t\sigma \in L(BPS(S))$.
    \item $\xi_{bps}(q_{bps}^{init}, t) \neq q^{dump}$. Clearly, $\xi_{bps}(q_{bps}^{init}, t_{1}\gamma)$ is a reaction state. According to Case 2 and Case 4 of $\xi_{bps}$ of $BPS(S)$, we have $\xi_{bps}(q_{bps}^{init}, t_{1}\gamma) = \xi_{bps}(q_{bps}^{init}, t_{1}\gamma t_{2})$, which is still a reaction state. In addition, by construction, any event in $\Sigma$ is defined at any reaction state, we have $t\sigma = t_{1}\gamma t_{2} \sigma \in L(BPS(S))$.
\end{enumerate}
Based on the above analysis, in any case, we have $t\sigma \in L(BPS(S))$, which completes the proof. \hfill $\blacksquare$ 

\section{Proof of Proposition V.2} 
\label{appendix: Proposition V.2} 
Since $L(G||S) \neq L(G||S')$, based on \textbf{Lemma V.1}, we have $L(B) = L(\mathscr{P}_{\Sigma_{o}}(G||S)) \neq L(\mathscr{P}_{\Sigma_{o}}(G||S')) = L(B')$, where $B' = \mathscr{P}_{\Sigma_{o}}(G||S') = (Q_{b'}, \Sigma, \xi_{b'}, q_{b'}^{init})$. Then we know that there exists $t \in \Sigma_{o}^{*} \cap L(B) \cap L(B')$ such that for any $i \in [0: |t|-1]$, the following conditions are satisfied:
\begin{enumerate}[C1)]
\setlength{\itemsep}{3pt}
\setlength{\parsep}{0pt}
\setlength{\parskip}{0pt}
    \item $En_{B}(\xi_{b}(q_{b}^{init}, \mathcal{P}_{i}(t))) \subseteq En_{S'}(\xi_{s'}(q_{s'}^{init}, \mathcal{P}_{i}(t)))$
    \item $(\forall (q_{g}, q_{s}) \in \xi_{b}(q_{b}^{init}, \mathcal{P}_{i}(t)))En_{G}(q_{g}) \cap En_{S'}(\xi_{s'}(q_{s'}^{init}, \\ \mathcal{P}_{i}(t))) \subseteq En_{B}(\xi_{b}(q_{b}^{init}, \mathcal{P}_{i}(t)))$
    \item $En_{B}(\xi_{b}(q_{b}^{init}, t)) \not\subseteq En_{S'}(\xi_{s'}(q_{s'}^{init}, t)) \vee \\ (\exists (q_{g}, q_{s}) \in \xi_{b}(q_{b}^{init}, t))En_{G}(q_{g}) \cap En_{S'}(\xi_{s'}(q_{s'}^{init}, t)) \\ \not\subseteq En_{B}(\xi_{b}(q_{b}^{init}, t))$
\end{enumerate}
Next, we consider two strings $u = \gamma_{0}t[1]\gamma_{1}\dots t[|t|-1]\gamma_{|t|-1}t[|t|]$ ($u = \varepsilon$ if $t = \varepsilon$) and $u\gamma_{|t|}$, where for any $i \in [0: |t|]$, we have $\gamma_{i} = En_{S'}(\xi_{s'}(q_{s'}^{init}, \mathcal{P}_{i}(t)))$. Since $t \in \Sigma_{o}^{*} \cap L(B) \cap L(B')$, we know that $t \in L(S')$. Thus, for any $j \in [1: |t|]$, it holds that $t[j] \in \gamma_{j-1}$. By construction, we have $u \in L(BT(S'))$. Next, we prove that $u \in L(BPS(S))$ by mathematical induction. For convenience, we denote $u = c_{1}\dots c_{|t|}$, where $c_{i} = \gamma_{i-1}t[i]$. The base case is to prove $c_{1} = \gamma_{0}t[1] \in L(BPS(S))$. If $t = \varepsilon$, then $u = \varepsilon$, which means that $c_{1} = \varepsilon \in L(BPS(S))$. Next, we only consider $t \neq \varepsilon$. Since $t \in L(\mathscr{P}_{\Sigma_{o}}(G||S))$, we have $t[1] \in L(\mathscr{P}_{\Sigma_{o}}(G||S))$. In addition, since the condition C1) and C2) hold, we know that for Case 1 of $\xi_{bps}$ of $BPS(S)$, the condition $\mathcal{C}_{1}$ and $\mathcal{C}_{2}$ are satisfied for $\gamma_{0}$ at the state $(q_{b}^{init})^{com}$ in $BPS(S)$. Thus, $\gamma_{0}t[1] \in L(BPS(S))$ and the base case holds. The induction hypothesis is $c_{1}\dots c_{k} = \gamma_{0}t[1]\gamma_{1}\dots \gamma_{k-1}t[k] \in L(BPS(S))$ and we need to prove $c_{1}\dots c_{k+1} = \gamma_{0}t[1]\gamma_{1}\dots \gamma_{k-1}t[k]\gamma_{k}t[k+1] \in L(BPS(S))$, where the hypothesis holds for $k \leq |t|-2$. It can be checked that $BPS(S)$ would transit to the state $(\xi_{b}(q_{b}^{init}, t[1]\dots t[k]))^{com}$ via the string $c_{1}\dots c_{k}$. Thus, we need to check whether the condition $\mathcal{C}_{1}$ and $\mathcal{C}_{2}$ in Case 1 of $\xi_{bps}$ of $BPS(S)$ are satisfied for $\gamma_{k}$ at the state $(\xi_{b}(q_{b}^{init}, t[1]\dots t[k]))^{com}$. 
Clearly, these two conditions hold as it is a special case of C1) and C2) when $i = k$. Thus, $u \in L(BPS(S))$. Since $BPNS(S) = BPS(S)||CE$ and $t[j] \in \gamma_{j-1}$ ($j \in [1: |t|]$), we know that $u \in L(BPNS(S))$.

Finally, we prove that $u\gamma_{|t|} \in L(BT(S'))$ and $u\gamma_{|t|} \notin L(BPS(S))$. By construction, $u\gamma_{|t|} \in L(BT(S'))$. Since the condition C3) holds, we know that for $BPS(S)$, at the state $q^{com} = (\xi_{b}(q_{b}^{init}, t))^{com} = \xi_{bps}(q_{bps}^{init}, u)$, it holds that either $En_{B}(q) \not\subseteq En_{S'}(\xi_{s'}(q_{s'}^{init}, t)) = \gamma_{|t|}$ or $(\exists (q_{g}, q_{s}) \in q)En_{G}(q_{g}) \cap En_{S'}(\xi_{s'}(q_{s'}^{init}, t)) = En_{G}(q_{g}) \cap \gamma_{|t|} \not\subseteq En_{B}(q)$, i.e., the conditions $\mathcal{C}_{1}$ and $\mathcal{C}_{2}$ in Case 1 of $\xi_{bps}$ of $BPS(S)$ are not satisfied for the state $q^{com} = \xi_{bps}(q_{bps}^{init}, u)$, rendering that $u\gamma_{|t|} \notin L(BPS(S))$ and thus $u\gamma_{|t|} \notin L(BPNS(S))$, which completes the proof.  \hfill $\blacksquare$

\section{Proof of Theorem V.1} 
\label{appendix: Theorem V.1} 
Based on \textbf{Proposition V.1}, we have LHS $\subseteq$ RHS. Next, we prove RHS $\subseteq$ LHS. Thus, we need to show that for any $t \in L(BPNS(S))$, we have $t \in$ LHS. We adopt the contradiction and assume that $t \notin$ LHS. Thus, we know that there exists a supervisor $\hat{S}$ such that $L(G||S) \neq L(G||\hat{S})$ and $t \in L(BT(\hat{S})) -$ LHS. Then, without loss of generality, we know that there exists $u \leq t$ such that $u = \gamma_{0}t_{1}\gamma_{1}\dots t_{m}\gamma_{m}$, where $m \in \mathbb{N}$ ($u = \gamma_{0}$ when $m = 0$) and the following conditions are satisfied:
\begin{enumerate}[1.]
\setlength{\itemsep}{3pt}
\setlength{\parsep}{0pt}
\setlength{\parskip}{0pt}
    \item $(\forall i \in [1:m])t_{i} \in (\gamma_{i-1} \cap \Sigma_{uo})^{*}(\gamma_{i-1} \cap \Sigma_{o})$ for $m \geq 1$. For convenience, we denote $t^{obs} = t_{1}^{\downarrow}\dots t_{m}^{\downarrow}$ for $m \geq 1$, and $t^{obs} = \varepsilon$ for $m = 0$.
    \item $(\forall i \in [1:m])En_{B}(\xi_{b}(q_{b}^{init}, \mathcal{P}_{i-1}(t^{obs}))) \subseteq \gamma_{i-1}$ for $m \geq 1$.
    \item $(\forall i \in [1:m])(\forall (q_{g}, q_{s}) \in \xi_{b}(q_{b}^{init}, \mathcal{P}_{i-1}(t^{obs})))En_{G}(q_{g})\\ \cap \gamma_{i-1} \subseteq En_{B}(\xi_{b}(q_{b}^{init}, \mathcal{P}_{i-1}(t^{obs})))$ for $m \geq 1$.
    \item $En_{B}(\xi_{b}(q_{b}^{init}, t^{obs})) \not\subseteq \gamma_{m} \vee (\exists (q_{g}, q_{s}) \in \xi_{b}(q_{b}^{init}, t^{obs}))\\En_{G}(q_{g}) \cap \gamma_{m} \not\subseteq En_{B}(\xi_{b}(q_{b}^{init}, t^{obs}))$
\end{enumerate}
Thus, $\mathcal{C}_{1} \wedge \mathcal{C}_{2}$ is not satisfied for the command $\gamma_{m}$ at the state $(\xi_{b}(q_{b}^{init}, t^{obs}))^{com}$ in Case 1 of $\xi_{bps}$ of $BPS(S)$. Thus, $u \notin L(BPS(S))$ and $t \notin L(BPS(S))$, which causes the contradiction. Hence, $t \in$ LHS, which completes the proof. \hfill $\blacksquare$

\section{Proof of Proposition VI.1} 
\label{appendix: Proposition VI.1} 
We denote $BT(S') = (Q_{bs'}, \Sigma \cup \Gamma, \xi_{bs'}, q_{bs'}^{init})$ and $BT(S')^{A} = (Q_{bs'}^{a}, \Sigma \cup \Gamma, \xi_{bs'}^{a}, q_{bs'}^{a,init})$. We shall show that $BT(S')^{A}$ is simulated by $BPNS^{A}(S)$. Let $R \subseteq Q_{bs'}^{a} \times Q_{bpns}^{a} = (Q_{bs'} \cup \{q^{detect}\}) \times (Q_{bpns} \cup \{q_{bpns}^{detect}\})$ be a relation defined such that 1) for any $q_{1} \in Q_{bs'} \subseteq Q_{bs'}^{a}$, any $q_{2} \in Q_{bpns} \subseteq Q_{bpns}^{a}$ and any $t \in L(BT(S')) \subseteq L(BPNS(S))$ such that $\xi_{bs'}(q_{bs'}^{init}, t) = q_{1}$ and $\xi_{bpns}(q_{bpns}^{init}, t) = q_{2}$, $(q_{1}, q_{2}) \in R$, and 2) $(q^{detect}, q_{bpns}^{detect}) \in R$.
We observe that, by construction, $(q_{bs'}^{a,init}, q_{bpns}^{a,init}) \in R$. Next, without loss of generality, we consider two states $q_{1} \in Q_{bs'}$ and $q_{2} \in Q_{bpns}$ such that $(q_{1}, q_{2}) \in R$. According to the definition of $R$, we know that there exists $t \in L(BT(S')) \subseteq L(BPNS(S))$ such that $\xi_{bs'}(q_{bs'}^{init}, t) = q_{1}$ and $\xi_{bpns}(q_{bpns}^{init}, t) = q_{2}$. Since $L(BT(S')) \subseteq \overline{(\Gamma\Sigma_{uo}^{*}\Sigma_{o})^{*}}$, there are three cases:
\begin{enumerate}[1.]
\setlength{\itemsep}{3pt}
\setlength{\parsep}{0pt}
\setlength{\parskip}{0pt}
    \item $t \in (\Gamma\Sigma_{uo}^{*}\Sigma_{o})^{*}\Gamma$. We know that $q_{1}$ and $q_{2}$ are reaction states, where only events in $\Sigma$ are defined. Then, for any $\sigma \in \Sigma$ such that $\xi_{bs'}^{a}(q_{1}, \sigma) = \hat{q}_{1}$, by construction, we have $\xi_{bpns}^{a}(q_{2}, \sigma)!$ and we denote $\xi_{bpns}^{a}(q_{2}, \sigma) = \hat{q}_{2}$. If $\sigma \in En_{BT(S')}(q_{1})$, then we have $\xi_{bs'}(q_{bs'}^{init}, t\sigma) = \hat{q}_{1}$ and $\xi_{bpns}(q_{bpns}^{init}, t\sigma) = \hat{q}_{2}$, i.e., $(\hat{q}_{1}, \hat{q}_{2}) \in R$. If $\sigma \notin En_{BT(S')}(q_{1})$, then we have the following two subcases. 1) $\sigma \in \Sigma_{uo}$. Since unobservable events are self-loop transitions, we know that $\hat{q}_{1} = q_{1}$ and $\hat{q}_{2} = q_{2}$, i.e., $(\hat{q}_{1}, \hat{q}_{2}) \in R$. 2) $\sigma \in \Sigma_{o}$. Then we know that $\hat{q}_{1} = q^{detect}$ and $\hat{q}_{2} = q_{bpns}^{detect}$. In addition, since $(q^{detect}, q_{bpns}^{detect}) \in R$, we still have $(\hat{q}_{1}, \hat{q}_{2}) \in R$.
    \item $t \in (\Gamma\Sigma_{uo}^{*}\Sigma_{o})^{*}\Gamma\Sigma_{uo}^{*}$. Since unobservable events are self-loops, this case can be reduced to Case 1.
    \item $t \in (\Gamma\Sigma_{uo}^{*}\Sigma_{o})^{*}$. We know that $q_{1}$ and $q_{2}$ are control states, where only events in $\Gamma$ are defined. For any $\gamma \in \Gamma$ such that $\xi_{bs'}^{a}(q_{1}, \gamma) = \hat{q}_{1}$, we have $t\gamma \in L(BT(S')) \subseteq L(BPNS(S)) \subseteq L(BPNS^{A}(S))$, i.e., $\xi_{bpns}^{a}(q_{2}, \sigma)!$ and we denote $\xi_{bpns}^{a}(q_{2}, \sigma) = \hat{q}_{2}$. Clearly, $(\hat{q}_{1}, \hat{q}_{2}) \in R$ as $\xi_{bs'}(q_{bs'}^{init}, t\gamma) = \hat{q}_{1}$ and $\xi_{bpns}(q_{bpns}^{init}, t\gamma) = \hat{q}_{2}$.
\end{enumerate}
Thus, $BT(S')^{A}$ is simulated by $BPNS^{A}(S)$, which completes the proof. \hfill $\blacksquare$

\section{Proof of Theorem VI.1} 
\label{appendix: Theorem VI.1} 
Based on \textbf{Proposition VI.1}, we have LHS $\subseteq$ RHS. Next, we prove RHS $\subseteq$ LHS, that is, for any $t \in$ RHS, we need to show $t \in$ LHS. Then there are two cases:
\begin{enumerate}[1.]
\setlength{\itemsep}{3pt}
\setlength{\parsep}{0pt}
\setlength{\parskip}{0pt}
    \item $t \in L(BPNS(S))$. Based on \textbf{Theorem V.1}, we have $t \in \bigcup\limits_{S' \in \mathscr{S}_{e}(S)}L(BT(S'))$. Since the contrscution of $BT(S')^{A}$ does not remove any transition defined in $BT(S)$, we have $t \in $ LHS.
    \item $t \notin L(BPNS(S))$ but $t \in L(BPNS^{A}(S))$. Then we need to prove $t \in$ LHS, i.e., for any $n \in [0: |t|]$, we have $\mathcal{P}_{n}(t) \in$ LHS.
    We adopt the mathematical induction. For the base case, it clearly holds as $\mathcal{P}_{0}(t) = \varepsilon \in$ LHS. The induction hypothesis is $\mathcal{P}_{k}(t) \in$ LHS, where the hypothesis holds for $k \leq |t|-2$, and we need to prove $\mathcal{P}_{k+1}(t) := \mathcal{P}_{k}(t)\sigma \in$ LHS. Then there are two subcases:
    \begin{enumerate}[a.]
    \setlength{\itemsep}{3pt}
    \setlength{\parsep}{0pt}
    \setlength{\parskip}{0pt}
        \item $\mathcal{P}_{k}(t) = t_{1}\gamma t_{2}$, where $t_{1} \in (\Gamma\Sigma_{uo}^{*}\Sigma_{o})^{*}$, $\gamma \in \Gamma$, $t_{2} \in (\gamma \cap \Sigma_{uo})^{*}$. Since $\mathcal{P}_{k}(t) \in$ LHS, there exists a supervisor $S' \in \mathscr{S}_{e}(S)$ such that $\mathcal{P}_{k}(t) \in L(BT(S')^{A})$. We denote $BT(S')^{A} = (Q_{bs'}^{a}, \Sigma \cup \Gamma, \xi_{bs'}^{a}, q_{bs'}^{a,init})$. By construction, we have $En_{BPNS^{A}(S)}(\xi_{bpns}^{a}(q_{bpns}^{a,init}, \mathcal{P}_{k}(t))) = En_{BT(S')^{A}}(\xi_{bs'}^{a}(q_{bs'}^{a,init}, \mathcal{P}_{k}(t)))$. Thus, $\mathcal{P}_{k+1}(t) = \mathcal{P}_{k}(t)\sigma \in L(BT(S')^{A}) \subseteq$ LHS.
        \item $\mathcal{P}_{k}(t) \in (\Gamma\Sigma_{uo}^{*}\Sigma_{o})^{*}$. In this case, $\sigma \in \Gamma$ because $L(BPNS^{A}(S)) \subseteq \overline{(\Gamma\Sigma_{uo}^{*}\Sigma_{o})^{*}}$. Since the construction of $BPNS^{A}(S)$ does not remove from $BPNS(S)$ any transition that is labelled by an event in $\Gamma$ and $\mathcal{P}_{k}(t) \in$ LHS, based on \textbf{Theorem V.1}, there exists a supervisor $S' \in \mathscr{S}_{e}(S)$ such that $\mathcal{P}_{k+1}(t) = \mathcal{P}_{k}(t)\sigma \in L(BT(S')^{A}) \subseteq$ LHS.
    \end{enumerate}    
\end{enumerate}
Thus, we have $t \in$ LHS, which completes the proof. \hfill $\blacksquare$ 

\section{Proof of Proposition VI.2} 
\label{appendix: Proposition VI.2} 
We adopt the contradiction and assume that $L(G||CE^{A}||BT(S')^{A}||\mathcal{A}) \not\subseteq L(G||CE^{A}||BPNS^{A}(S)||\hat{\mathcal{A}})$. Thus, there exists $t \in \overline{(\Gamma\Sigma_{uo}^{*}\Sigma_{o})^{*}}$ such that $t \in L(G||CE^{A}||BT(S')^{A}||\mathcal{A})$ and $t \notin L(G||CE^{A}||BPNS^{A}(S)||\hat{\mathcal{A}})$. Based on \textbf{Proposition VI.1}, we have $t \in L(G||CE^{A}||BPNS^{A}(S))$ and $t \notin L(\hat{\mathcal{A}})$. Then we have the following two cases:
\begin{enumerate}[1.]
\setlength{\itemsep}{3pt}
\setlength{\parsep}{0pt}
\setlength{\parskip}{0pt}
    \item $t \notin L(\mathcal{P}_{r})$. By construction, $G||CE^{A}||BT(S')^{A}||\mathcal{A}$ would reach the state $(q, q_{ce}^{a}, q_{bs'}^{a}, q_{a})$, where $q_{bs'}^{a} = q^{detect}$ via the string $t$, which causes the contradiction as $\mathcal{A}$ is covert.
    \item $t \in L(\mathcal{P}_{r})$. Since $t \notin L(\hat{\mathcal{A}})$, we know that there exists $\hat{t} = \hat{t}'\gamma\sigma_{1}\sigma_{2}\dots\sigma_{n}\sigma_{o} \in L(\mathcal{P})$ such that: 1) $(\exists u \leq t)P_{\Sigma_{o,a} \cup \Gamma}(\hat{t}) = P_{\Sigma_{o,a} \cup \Gamma}(u)$, where $P_{\Sigma_{o,a} \cup \Gamma}: (\Sigma \cup \Gamma)^{*} \rightarrow (\Sigma_{o,a} \cup \Gamma)^{*}$, 2) $\xi_{\mathcal{P}}(q_{\mathcal{P}}^{init}, \hat{t}) \in Q_{bad}$, 3) $\hat{t}^{\downarrow} = \sigma_{o} \in (\Sigma_{c,a} - \gamma) \cap \Sigma_{o}$, and 4) $(\forall i \in [1:n]) \sigma_{i} \in (\gamma \cup \Sigma_{c,a}) \cap \Sigma_{uo}$. Since $P_{\Sigma_{o,a} \cup \Gamma}(\hat{t}) = P_{\Sigma_{o,a} \cup \Gamma}(u)$, we know that $u = t'\gamma\sigma_{1}'\sigma_{2}'\dots\sigma_{m}'\sigma_{o}$ such that: 1) $P_{\Sigma_{o,a} \cup \Gamma}(t') = P_{\Sigma_{o,a} \cup \Gamma}(\hat{t}')$, and 2) $(\forall i \in [1:m]) \sigma_{i}' \in (\gamma \cup \Sigma_{c,a}) \cap \Sigma_{uo}$.
    Since $\sigma_{o} \in (\Sigma_{c,a} - \gamma) \cap \Sigma_{o}$, we know that $BPNS^{A}(S)$ would reach the state $q_{bpns}^{detect}$ via the string $u$, implying that $t = u$. Thus, $\xi_{\mathcal{P}}(q_{\mathcal{P}}^{init}, t) \in Q_{bad}$, which causes the contradiction with the fact that $t \in L(\mathcal{P}_{r})$.
\end{enumerate}
Henceforth, the assumption 
does not hold and the proof is completed. 
\hfill $\blacksquare$


\section{Proof of Theorem VI.2} 
\label{appendix: Theorem VI.2} 
Based on \textbf{Corollary VI.1}, we have RHS $\subseteq$ LHS. Next, we show LHS $\subseteq$ RHS. We adopt the contradiction and assume LHS $\not\subseteq$ RHS. Then we know that there exists $\varepsilon \neq t \in \overline{(\Gamma\Sigma_{uo}^{*}\Sigma_{o})^{*}}$ such that $t \in$ LHS and $t \notin$ RHS. Hence, $t$ can be executed in $G$, $CE^{A}$, $BPNS^{A}(S)$ and $\hat{\mathcal{A}}$, after we lift their alphabets to $\Sigma \cup \Gamma$, and $G$ reaches the state in $Q_{d}$ via $t$. Then, based on \textbf{Theorem VI.1}, we know that there exists $S' \in \mathscr{S}_{e}(S)$ such that $t \in L(BT(S')^{A})$. Thus, $t \in L_{m}(G||CE^{A}||BT(S')^{A}||\hat{\mathcal{A}})$, i.e., $\hat{\mathcal{A}}$ is damage-reachable against $S'$. In addition, $\hat{\mathcal{A}}$ is covert against $S'$; otherwise, there exists $t' \in L(G||CE^{A}||BT(S')^{A}||\hat{\mathcal{A}})$ such that $BT(S')^{A}$ reaches the state $q^{detect}$ via the string $t'$, which results in that $L(G||CE^{A}||BPNS^{A}(S)||\hat{\mathcal{A}})$ reaches some state in $Q_{bad}$ via the string $t'$ and the contradiction is caused. Thus, $\hat{\mathcal{A}}$ is covert and damage-reachable against $S'$ and we have $\hat{\mathcal{A}} \in \mathscr{A}(S')$, which means that $t \in$ RHS and this causes the contradiction. Hence, LHS $\subseteq$ RHS, and the proof is completed. \hfill $\blacksquare$

\section{Proof of Proposition VI.3} 
\label{appendix: Proposition VI.3} 
Firstly, we prove $L(S_{0}) \subseteq L(BPNS(S))$. We adopt the contradiction and assume that $L(S_{0}) \not\subseteq L(BPNS(S))$. Since $L(S_{0}) \subseteq \overline{(\Gamma\Sigma_{uo}^{*}\Sigma_{o})^{*}}$ and $L(BPNS(S)) \subseteq \overline{(\Gamma\Sigma_{uo}^{*}\Sigma_{o})^{*}}$, we have the following two cases. 1) There exists $t \in L(S_{0}) \cap L(BPNS(S))$ and $\gamma \in \Gamma$ such that $t\gamma \in L(S_{0})$ and $t\gamma \notin L(BPNS(S))$. Thus, $t\gamma \in L(S_{0}) \subseteq L(S_{0}^{A}) \subseteq L(BPNS^{A}(S))$. Since the construction of $BPNS^{A}(S)$ from $BPNS(S)$ does not add any transition labelled by a control command, we have $t\gamma \in L(BPNS(S))$, which causes the contradiction. 2) There exists $t \in (\Sigma \cup \Gamma)^*$, $\gamma \in \Gamma$, $t' \in (\gamma \cap \Sigma_{uo})^{*}$ and $\sigma \in \Sigma$ such that $t\gamma t' \in L(S_{0}) \cap L(BPNS(S))$, $t\gamma t'\sigma \in L(S_{0})$ and $t\gamma t'\sigma \notin L(BPNS(S))$. Thus, $t\gamma t'\sigma \in L(S_{0}) \subseteq L(S_{0}^{A}) \subseteq L(BPNS^{A}(S))$. Based on the construction of $BPNS^{A}(S)$ from $BPNS(S)$, we have $\sigma \in \Sigma - \gamma$. However, this would violate the structure of $S_{0}$, which causes the contradiction. 
Thus, $L(FNS(S)) \subseteq L(S_{0}) \subseteq L(BPNS(S))$. \hfill $\blacksquare$

\section{Proof of Theorem VI.3} 
\label{appendix: Theorem VI.3} 
Firstly, we prove LHS $\subseteq$ RHS. Thus, we shall show for any $S' \in \mathscr{S}_{f}(S)$, we have $L(BT(S')) \subseteq L(FNS(S))$. Based on \textbf{Proposition VI.1}, we have $L(BT(S')^{A}) \subseteq L(BPNS^{A}(S))$. Next, we prove $L(BT(S')^{A}) \subseteq L(S_{0}^{A})$. We adopt the contradiction and assume that $L(BT(S')^{A}) \not\subseteq L(S_{0}^{A})$. By synthesis, we know there exists $u \in L(BPNS^{A}(S))$, $\gamma \in \Gamma$ and $v \in (\gamma \cup \Sigma_{c,a})^{*} - \{\varepsilon\}$ such that 
\[
\begin{aligned}
u\gamma v \in L_{m}(\mathcal{P}) \wedge [(&\exists t \in L(BT(S')^{A} ))t\gamma \in L(BT(S')^{A}) \wedge \\ &t\gamma \notin L(S_{0}^{A}) \wedge P_{\Sigma_{o} \cup \Gamma}(t) = P_{\Sigma_{o} \cup \Gamma}(u)]
\end{aligned}
\]
By construction, we have $u\gamma v \in L(BT(S')^{A})$, i.e., there exists a covert and damage-reachable actuator attacker against $S'$, which is contradictory to the fact that $S'$ is resilient. Thus, $L(BT(S')^{A}) \subseteq L(S_{0}^{A})$. Then, it can be checked that $L(BT(S')) \subseteq L(S_{0})$. Next, we prove $L(BT(S')) \subseteq L(FNS(S))$. We adopt the contradiction and assume that $L(BT(S')) \not\subseteq L(FNS(S))$. Then, according to \textbf{Procedure 3}, without loss of generality, we know that there exists $k \geq 0$ such that $L(BT(S')) \subseteq L(S_{k})$ and $L(BT(S')) \not\subseteq L(S_{k+1})$. Then we know that there exists $u' \in L(S_{k})$, $\gamma'' \in \Gamma$ and $\sigma' \in \gamma \cap \Sigma_{o}$ such that 
\[
\begin{aligned}
&En_{S_{k}}(\xi_{S_{k}}(q_{S_{k}}^{init}, u'\gamma'' \sigma')) = \varnothing \wedge
\\& [(\exists t' \in L(BT(S'))) t'\gamma'' \in L(BT(S')) \wedge t'\gamma'' \notin L(S_{k+1}) \wedge \\& P_{\Sigma_{o} \cup \Gamma}(u') = P_{\Sigma_{o} \cup \Gamma}(t')]
\end{aligned}
\]
Similarly, we have $u'\gamma'' \in L(BT(S'))$, and thus $u'\gamma'' \sigma' \in L(BT(S'))$. Since $L(BT(S')) \subseteq L(S_{k})$ and there is always a control command in $\Gamma$ defined at any control state of $BT(S')$, we have $En_{S_{k}}(\xi_{S_{k}}(q_{S_{k}}^{init}, u'\gamma'' \sigma')) \neq \varnothing$, which causes the contradiction. Thus, $L(BT(S')) \subseteq L(FNS(S))$.

Secondly, we prove RHS $\subseteq$ LHS. Thus, we need to show for any $t \in$ RHS, we have $t \in$ LHS. Firstly, we generate an automaton $T$ such that $L_{m}(T) = t$. Then we compute its subset construction $\mathscr{P}_{\Sigma_{o} \cup \Gamma}(T) = (Q_{t}, \Sigma \cup \Gamma, \xi_{t}, q_{t}^{init})$. By construction, we could denote $Q_{t} = Q_{t}^{rea} \dot{\cup } Q_{t}^{com}$, where $Q_{t}^{rea}$ is the set of reaction states and $Q_{t}^{com}$ is the set of control states. Then we construct a new automaton $NC = (Q_{nc}, \Sigma_{nc}, \xi_{nc}, q_{nc}^{init})$, where $Q_{nc} = Q_{t} \cup \{q^{obs}\} \cup \{q^{\gamma}|\gamma \in \Gamma\}$, $q_{nc}^{init} = q_{t}^{init}$, $\Sigma_{nc} = \Sigma \cup \Gamma$, and $\xi_{nc}$ is defined as follows:
\begin{enumerate}[1.]
\setlength{\itemsep}{3pt}
\setlength{\parsep}{0pt}
\setlength{\parskip}{0pt}
    \item $(\forall q, q' \in Q_{t})(\forall \sigma \in \Sigma \cup \Gamma)\xi_{t}(q, \sigma) = q' \Rightarrow \xi_{nc}(q, \sigma) = q'$
    \item $(\forall q \in Q_{t}^{rea})(\forall \sigma \in \Sigma_{uo})\neg \xi_{t}(q, \sigma)! \Rightarrow \xi_{nc}(q, \sigma) = q$
    \item $(\forall q \in Q_{t}^{rea})(\forall \sigma \in \Sigma_{o})\neg \xi_{t}(q, \sigma)! \Rightarrow \xi_{nc}(q, \sigma) = q^{obs}$
    \item $(\forall q \in Q_{t}^{com})En_{\mathscr{P}_{\Sigma_{o} \cup \Gamma}(T)}(q) = \varnothing \Rightarrow (\forall \gamma \in \Gamma)\xi_{nc}(q, \gamma) = q^{\gamma}$
    \item $(\forall \gamma \in \Gamma)\xi_{nc}(q^{obs}, \gamma) = q^{\gamma}$
    \item $(\forall \gamma \in \Gamma)(\forall \sigma \in \gamma \cap \Sigma_{o}) \xi_{ce}(q^{\gamma}, \sigma) = q^{obs}$.
    \item $(\forall \gamma \in \Gamma)(\forall \sigma \in \gamma \cap \Sigma_{uo}) \xi_{ce}(q^{\gamma}, \sigma) = q^{\gamma}$. 
\end{enumerate}
Then we compute $NCS = NC||FNS(S) = (Q_{ncs}, \Sigma \cup \Gamma, \xi_{ncs}, q_{ncs}^{init})$. By construction, we denote $Q_{ncs} = Q_{ncs}^{rea} \dot{\cup} Q_{ncs}^{com}$, where $Q_{ncs}^{rea}$ is the set of reaction states and $Q_{ncs}^{com}$ is the set of control states. Based on $NCS$, we generate a bipartite supervisor, denoted as $BT = (Q_{bt}, \Sigma_{bt}, \xi_{bt}, q_{bt}^{init})$, where $Q_{bt} = Q_{ncs} = Q_{ncs}^{rea} \dot{\cup} Q_{ncs}^{com}$, $q_{bt}^{init} = q_{ncs}^{init}$, $\Sigma_{bt} = \Sigma \cup \Gamma$, and $\xi_{bt}$ is defined as follows:
\begin{enumerate}[1.]
\setlength{\itemsep}{3pt}
\setlength{\parsep}{0pt}
\setlength{\parskip}{0pt}
    \item $(\forall q, q' \in Q_{bt})(\forall \sigma \in \Sigma)\xi_{ncs}(q, \sigma) = q' \Rightarrow \xi_{bt}(q, \sigma) = q'$
    \item For any control state $q \in Q_{ncs}^{com}$, we randomly pick a control command $\gamma \in En_{NCS}(q)$ and define that: for any reaction state $q' \in Q_{ncs}^{rea}$, if $\xi_{ncs}(q, \gamma) = q'$, then $\xi_{bt}(q, \gamma) = q'$ and for any control command $\gamma' \in En_{NCS}(q) - \{\gamma\}$, we have $\neg\xi_{bt}(q, \gamma')!$.
\end{enumerate}
Finally, we generate the automaton $Ac(BT)$. For convenience, we shall still denote $Ac(BT)$ as $BT$. 
Next, we firstly prove $BT$ is control equivalent to $S$. We adopt the contradiction and assume that $BT$ is not control equivalent to $S$. Based on \textbf{Proposition V.2}, we have $L(BT) \not\subseteq L(BPNS(S))$. Since $BT = NC||FNS(S)$, we have $L(BT) \subseteq L(FNS(S))$. Based on \textbf{Proposition VI.3}, we have $L(BT) \subseteq L(FNS(S)) \subseteq L(BPNS(S))$, which causes the contradiction. Hence, $BT$ is control equivalent to $S$.
Secondly, we prove $BT$ is resilient. We adopt the contradiction and assume $BT$ is not resilient. We denote the version of $BT$ under attack as $BT^{A}$. Clearly, we have $L(BT^{A}) \subseteq L(S_{0}^{A})$. Since $BT$ is not resilient, we know that there exists an attacker $\mathcal{A} \in \mathscr{A}(BT)$ and a string $t \in L(BT^{A}) \subseteq L(S_{0}^{A})$ such that $t \in L_{m}(G||CE^{A}||BT^{A}||\mathcal{A})$. Based on \textbf{Proposition VI.2}, we have $t \in L_{m}(\mathcal{P}) = L_{m}(G||CE^{A}||BPNS^{A}(S)||\hat{\mathcal{A}})$. By synthesis, we know that $t \notin L(S_{0}^{A})$, which causes the contradiction. Thus, $BT$ is resilient.
Finally, we show that $t \in L(BT)$. By construction, $t \in L(NC)$. Since $t \in$ RHS = $L(FNS(S))$ and $NCS = NC||FNS(S)$, we have $t \in L(NCS)$. We adopt the contradiction and assume that $t \notin L(BT)$. By construction, there exists $t' \leq t$ and $\gamma \in \Gamma$ such that 1) $t'\gamma \leq t$, 2) $|En_{NCS}(\xi_{ncs}(q_{ncs}^{init}, t'))| \geq 2$, and 3) we do not pick the control command $\gamma$ at the control state $\xi_{ncs}(q_{ncs}^{init}, t')$ when we construct $BT$. However, by construction, there is only one control command defined at the state $\xi_{ncs}(q_{ncs}^{init}, t')$, which causes the contradiction. Thus, $t \in L(BT)$.
Based on the above analysis, $BT \in \mathscr{S}_{f}(S)$, and $t \in L(BT) \subseteq$ LHS. Thus, RHS $\subseteq$ LHS, which completes the proof. \hfill $\blacksquare$

\section{Proof of Proposition VI.4} 
\label{appendix: Proposition VI.4} 
Firstly, at any reachable control state, only one control command in $\Gamma$ is defined, and such a transition would lead to a reaction state. Secondly, at any reachable reaction state, which is reached from a control state via a transition labelled by $\gamma \in \Gamma$, all the events in $\gamma$ are defined, where any event in $\gamma \cap \Sigma_{uo}$ is a self-loop transition and any event in $\gamma \cap \Sigma_{o}$ would lead to a control state. Thus, $FS(S)$ is consistent with a bipartite supervisor structure. The analysis of $FS(S)$ being control equivalent and resilient is similar to that of $BT$ in the proof of \textbf{Theorem VI.3}, which completes the proof. \hfill $\blacksquare$


\end{appendices}

\begin{IEEEbiography}[
{
\includegraphics[width=1.0in,height=1.40in,clip,keepaspectratio]{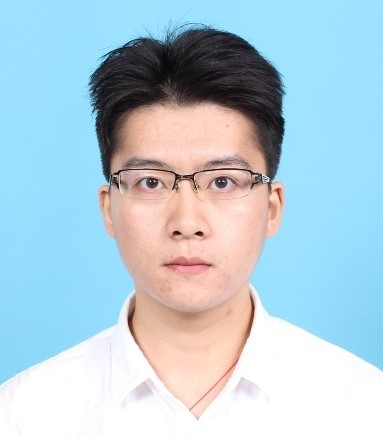}
}
]
{Ruochen Tai}
received the B.E. degree in electrical engineering and automation from the Nanjing University of Science and Technology in 2016, and M.S. degree in automation from the Shanghai Jiao Tong University in 2019. He is currently pursuing the Ph.D. degree with Nanyang Technological University, Singapore. His current research interests include cyber security, multi-robot systems, soft robotics, formal methods, and discrete-event systems.
\end{IEEEbiography}
\begin{IEEEbiography}[
{
\includegraphics[width=1.0in,height=1.40in,clip,keepaspectratio]{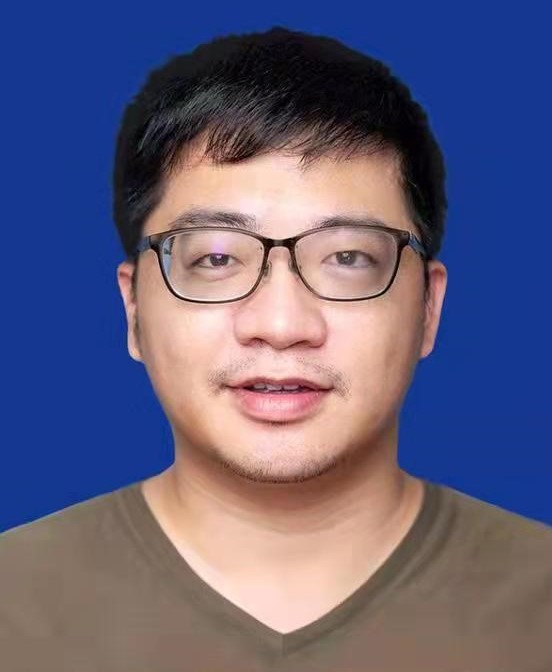}
}
]
{Liyong Lin}
received the B.E. degree and Ph.D. degree in electrical engineering in 2011 and 2016, respectively, both from Nanyang Technological University. From June 2016 to October 2017, he was a postdoctoral fellow at the University of Toronto. Since December 2017, he has been working as a research fellow at the Nanyang Technological University. His main research interests include supervisory control theory, formal methods, automated reasoning, flexible manufacturing and smart logistics. 
\end{IEEEbiography}
\begin{IEEEbiography}[
{
\includegraphics[width=1in,height=1.3in,clip,keepaspectratio]{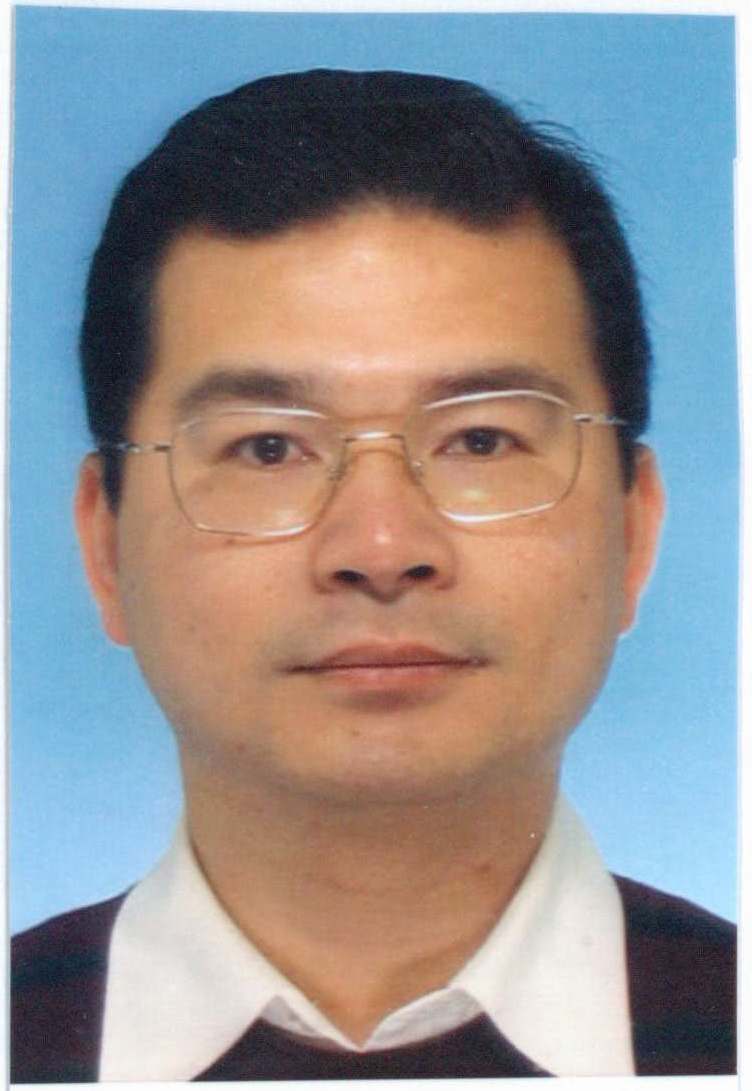}
}
]
{Rong Su} received the Bachelor of Engineering degree from University of Science and Technology of China in 1997, and the Master of Applied Science degree and PhD degree from University of Toronto, in 2000 and 2004, respectively. He was affiliated with University of Waterloo and Technical University of Eindhoven before he joined Nanyang Technological University in 2010. Currently, he is an associate professor in the School of Electrical and Electronic Engineering. Dr. Su's research interests include multi-agent systems, cybersecurity of discrete-event systems, supervisory control, model-based fault diagnosis, control and optimization in complex networked systems with applications in flexible manufacturing, intelligent transportation, human-robot interface, power management and green buildings. In the aforementioned areas he has more than 220 journal and conference publications, and 5 granted USA/Singapore patents. Dr. Su is a senior member of IEEE, and an associate editor for Automatica, Journal of Discrete Event Dynamic Systems: Theory and Applications, and Journal of Control and Decision. He was the chair of the Technical Committee on Smart Cities in the IEEE Control Systems Society in 2016-2019, and is currently the chair of IEEE Control Systems Chapter, Singapore.

\end{IEEEbiography}

\end{document}